\documentclass[11pt]{article}
\usepackage[margin=1in]{geometry}
\usepackage{amsmath,amssymb,mathtools}
\usepackage{amsthm}
\usepackage{authblk}
\usepackage[hidelinks]{hyperref}
\usepackage{tikz}
\usepackage{microtype}
\usepackage{placeins}
\usepackage{tabularx}

\newcommand{\dd}{\mathrm d}
\newcommand{\cH}{\mathcal H}
\newcommand{\cD}{\mathcal D}
\newcommand{\cR}{\mathcal R}

\newcommand{\cK}{\mathcal K}
\newcommand{\cQ}{\mathcal Q}
\newcommand{\cB}{\mathcal B}
\newcommand{\cC}{\mathcal C}

\newcommand{\loc}{\mathrm{loc}}
\newcommand{\dyn}{\mathrm{dyn}}

\newcommand{\matt}{\mathrm{matt}}

\newtheorem{theorem}{Theorem}
\newtheorem{proposition}[theorem]{Proposition}
\newtheorem{corollary}[theorem]{Corollary}

\theoremstyle{definition}
\newtheorem{definition}[theorem]{Definition}
\theoremstyle{remark}
\newtheorem{remark}{Remark}

\title{Dynamical entropy on local causal horizons:
the Einstein equation of state and canonical-energy flux}
\author{Roh-Suan Tung\thanks{Electronic address:
\href{mailto:rstung@cycu.edu.tw}{\texttt{rstung@cycu.edu.tw}}}}
\affil{Department of Physics and Center for Theoretical Physics,
Chung Yuan Christian University, Taoyuan 320, Taiwan}
\date{}

\begin{document}
\maketitle

\begin{abstract}
For a twist-free null hypersurface, the cut-local
Hollands--Wald--Zhang (HWZ) representative
\(S_{\loc}=(4G)^{-1}\int(1-V\theta)\dd A\) obeys an exact off-shell
Raychaudhuri identity containing no term linear in the instantaneous
expansion, so no final equilibrium cut is required.  Given an entropy
density \(1/(4G)\), the Unruh temperature \(\kappa/2\pi\) for an
approximate local boost, local equilibrium cuts, and stress-tensor
conservation, equating the leading equilibrium change of
\(S_{\loc}\) to matter boost heat yields \(R_{kk}=8\pi GT_{kk}\) for
every null direction, and hence the Einstein equation with an
undetermined cosmological constant.  This is an equation of state in
Jacobson's conditional sense: a consistency condition on these
inputs, not a derivation of them.

On a fixed caustic-free generator bundle with
\(\dd A=J\,\dd A_{\rm ref}\), the functional
\(S_{\rm L}=(4G)^{-1}\int(1+\ln J-V\theta)\dd A_{\rm ref}\) satisfies
an exact balance, normalized per reference cell, whose curvature,
shear, and expansion terms all enter with positive sign; the null
convergence condition then makes \(S_{\rm L}\) nondecreasing for
\(V\geq0\), until the first caustic.  Requiring that balance throughout an interval of
positive affine parameter, on every small bundle and in every null
direction, is conversely equivalent to the Einstein equation under
stress-tensor conservation and the stated normalization.
Cut-locality, positivity, and first-order HWZ normalization select the density \(1+\ln J-V\theta\)
uniquely within a natural two-function class.  Both functionals depend
on the affine origin.  We give their transformation laws under an
affine shift, identify the anchoring-invariant content, and show that
the origin is physical data fixed by the equilibrium cut.

The entropy interpretation is restricted accordingly.  Near an
equilibrium cut, \(S_{\rm L}\) agrees with the HWZ dynamical entropy at
first order, and the temperature-weighted change of its Hessian
across a horizon segment equals the HWZ modified canonical-energy
flux under the stated stationary-vacuum perturbation hypotheses.  At finite
amplitude, \(S_{\rm L}\) is established only as a monotone geometric
functional; reading it as an entropy requires an additional
coarse-graining postulate that is stated but not derived.
\end{abstract}


\section{Introduction}

Jacobson derived the Einstein equation by demanding that the
Clausius relation
\begin{equation}
 \delta Q=T_{\rm U}\,\delta S
 \label{eq:Clausius}
\end{equation}
hold for all local Rindler horizons through every spacetime point
\cite{Jacobson}.  The heat is the boost energy of matter, the
temperature is the Unruh temperature, and the entropy is proportional
to horizon area.  Raychaudhuri's equation converts the area response
into the null-null Ricci tensor.  Requiring
\eqref{eq:Clausius} for every null direction gives the Einstein
equation, with the cosmological constant left as an integration
constant.  A distinct modern route derives the field equation from
stationarity of vacuum entanglement entropy in small causal diamonds
\cite{JacobsonEE}; the present paper instead stays within the
causal-horizon Clausius framework and modifies its cut functional.

Jacobson's original argument is intrinsically a near-equilibrium,
first-order construction.  Two completion problems then arise.  The
first is perturbative: at second order the square of the linearized
shear enters Raychaudhuri's equation.  It is positive and has long
been interpreted as internal entropy production
\cite{ElingGuedensJacobson,ChircoLiberati}.  Hollands, Wald, and Zhang
gave this term a sharper meaning by constructing a dynamical
black-hole entropy whose vacuum second-order change is controlled by
a modified canonical-energy flux \cite{HWZ}.  In Einstein gravity
that flux is the shear square.

The second problem is genuinely nonlinear: can one obtain a positive
finite-amplitude balance without imposing a final equilibrium cut?
On every fixed, caustic-free bundle of null generators, the cut-local
HWZ representative differs by an exact convex area-Jacobian storage
term from a generator-relative functional \(S_{\rm L}\).
Raychaudhuri's equation gives \(S_{\rm L}\) a derivative in which
every term enters with positive sign, so that it is nonnegative
whenever \(R_{kk}\geq0\).  An important qualification is that its matter term is
normalized with the reference-generator measure rather than the
instantaneous physical area measure.  The exact identity is therefore
a generator-normalized balance, not by itself an ordinary
finite-amplitude Clausius law.  Its equilibrium Hessian is the
cut-renormalized HWZ entropy Hessian, so the second-order
canonical-energy law is the physically established equilibrium sector
of the nonlinear construction.

Second variations of the dynamical entropy have also been obtained
directly from Raychaudhuri's equation for stationary black-hole
horizons, including in Einstein-frame treatments of higher-curvature
theories \cite{KongTianZhangZhao}.  Related second laws on null
hypersurfaces have been obtained from Noether currents
\cite{RignonBret}.  Our distinct purpose is to place
that mechanism inside an off-shell local causal-horizon derivation of
the field equation, with an approximate boost and a pointwise
shrinking-patch limit.  Earlier Rindler-horizon mechanics instead
uses the Einstein equation and a final boundary condition to relate
area to matter flux \cite{BianchiSatz}.  The cut-local entropy
\eqref{eq:Sloc} (defined in Sec.~\ref{sec:offshell}) removes that
teleological step.

During completion of the present analysis, Fu, Ji, Zhou, and Cao
constructed a matter-sensitive modified canonical energy for
second-order perturbations of stationary bifurcate black-hole
horizons \cite{FuJiZhouCao}.  Their four-dimensional horizon
integrand contains the same shear, expansion, and area-weighted
stress-tensor terms found below.  The settings and logical roles are
different: their field equation is assumed and their goal is a
black-hole apparent-horizon area law, while our goal is an off-shell
local equation of state followed by its on-shell Killing-horizon
Hessian.  They have subsequently extended
their apparent-horizon and modified-canonical-energy analysis to
stationary electrovacuum backgrounds in Einstein--Maxwell theory
\cite{FuJiZhouCaoEM}.  That extension remains an on-shell global
black-hole construction, whereas the field equation in the present
work is obtained first from the local off-shell balance.
Furugori, Nishii, Yoshida, and Yoshimura have independently related
the linearized HWZ entropy to Wall entropy on an entropic marginally
outer trapped surface \cite{FurugoriEtAl}.  The second-order comparison
below instead isolates the precise quadratic obstruction to such an
identification for the cut-renormalized Einstein entropy.
Jia, Qi, and Gao have further argued, using replica methods, that the
HWZ dynamical entropy has an entanglement interpretation on the
perturbative apparent horizon and that the corresponding modified
matter entropy is the matter entanglement across that surface
\cite{JiaQiGao}.  Together with the HWZ relation between the quantum
null energy condition and the generalized second law, this supplies
an entropy interpretation near stationary equilibrium.  It does not
constitute a microscopic derivation of the finite-amplitude
generator-relative functional introduced below.

There is an apparent circularity that must be avoided.  Canonical
energy is normally defined from a gravitational Lagrangian
\cite{HollandsWald}, in the tradition of quasilocal charges built from
Hamiltonian boundary terms \cite{ChenNesterTung}, whereas
Jacobson's argument aims to derive the gravitational field equation.
Our organization is consequently strict:
\begin{enumerate}
 \item an exact entropy identity is derived off shell from null
       geometry alone;
 \item the leading equilibrium balance with matter heat yields the
       Einstein equation;
 \item the same off-shell geometry is completed by an exact
       finite-amplitude generator-relative identity;
 \item only afterward is the equilibrium Hessian restricted to the
       Einstein solution space and identified with the HWZ canonical
       flux.
\end{enumerate}
Neither the HWZ current nor the Einstein equation is used to obtain the
off-shell identities.  The canonical current is instead the
covariant-phase-space representation of the nonlinear law's
equilibrium Hessian.

The sense in which the first-order result is an ``equation of state''
should be stated at the outset.  We use the term exactly as Jacobson
did \cite{Jacobson}: the Einstein equation is obtained as the
condition under which a postulated local Clausius relation is
consistent for all local causal horizons.  The derivation rests on
explicit inputs, each of which is assumed rather than derived here:
(a) an entropy density \(1/(4G)\) and the cut-local entropy
\eqref{eq:Sloc}; (b) the Unruh temperature \(T_{\rm U}=\kappa/2\pi\)
assigned to the orbits of an approximate boost; (c) the leading
shrinking-patch Clausius balance itself; (d) conservation
\(\nabla^aT_{ab}=0\); (e) the existence of local equilibrium cuts
\eqref{eq:equilibrium}; and (f) the anchoring of the affine origin,
and hence of the boost, at the equilibrium cut
(Sec.~\ref{sec:affine}).  These inputs appear as hypotheses of
Theorem~\ref{thm:Einstein}.  The result is therefore a conditional
statement: thermodynamic input of this form is compatible only with
the Einstein equation.  It is not a derivation of gravitational
dynamics from microscopic statistical mechanics, and Newton's
constant is fixed by input (a) rather than predicted.

We also fix terminology, since the thermodynamic words carry more than
the mathematics establishes.  The words ``entropy'', ``heat'', and
``second law'' are used for the HWZ dynamical entropy and its
equilibrium variations, and for its cut-local representative
\(S_{\loc}\) by postulate (input (a)); in the equilibrium regime they are anchored by the HWZ
physical-process first law and canonical-energy positivity
\cite{HWZ}.  For the finite-amplitude functional \(S_{\rm L}\) we say
``generator-relative functional'' and ``monotonicity''; the word
``entropy'' is attached to it only conditionally, under the explicit
coarse-graining postulate of Sec.~\ref{sec:physicalentropy}.  Positive
quadratic terms are called ``fluxes'' (shear) or ``focusing''
(expansion); we do not call them entropy production, and we make no
\(H\)-theorem claim.

The local boost is another delicate point.  A generic curved
spacetime has no exact boost Killing field.  Null normal coordinates
nevertheless provide a vector field that agrees with
\(\kappa Vk^a\) on a selected generator, obeys the Killing identity
there, and fails to be Killing only quadratically in distance
\cite{GuedensNNC,GuedensJacobsonSarkar}.  We use only the pointwise
shrinking limit required by the equation-of-state argument.  A uniform
second-order finite-patch estimate would require additional
perturbation norms and boundary hypotheses and is not claimed here.

We use the metric signature \((-,+,+,+)\).  The principal results are as follows.
Theorem~\ref{thm:offshell} expresses the derivative of the cut-local
HWZ entropy in terms of the Raychaudhuri integrand alone, using no
field equation.  The cancellation due to \(-V\theta\) makes the
identity local in the cut and removes the final boundary condition
needed when area entropy is used by itself \cite{BianchiSatz}.

The finite-amplitude geometric result is
Theorem~\ref{thm:nonlinearidentity}.  The area-Jacobian functional
\(S_{\rm L}\) converts the full Raychaudhuri equation into the
sign-organized identity \eqref{eq:nonlinearoffshell}, which is
nonnegative under the null convergence condition; on Einstein
solutions the latter follows from the null energy condition.
Theorem~\ref{thm:balanceequivalence} shows conversely that requiring
the corresponding balance throughout an interval of positive affine
parameter, on every small bundle and in every null direction, is
equivalent to the Einstein equation, so the finite-amplitude identity
characterizes the field equation rather than presupposing it.  Equation
\eqref{eq:physicalgeneratorbalance} separates the instantaneous-area
boost-weighted flux from the reference-measure conversion needed by
the Lagrangian normalization.  The conversion is not a new energy
channel: its Hessian removes the area-strain part of flux through a
deforming cell.
The difference from the reference-cut HWZ representative is a convex
storage functional, while \eqref{eq:generalnonlinearidentity} proves
the structural role of the reference measure only within the stated
area-Jacobian endpoint class.  Theorem~\ref{thm:selection} sharpens
this: within the two-function class \(\Psi(J)+f(V)\theta\),
cut-locality and pointwise nonnegativity select the density
\(1+\ln J-V\theta\) uniquely up to normalization, and first-order HWZ
agreement fixes the normalization.  The shear-free solution
\eqref{eq:homogeneousfocusing} is an explicit warning against
overinterpretation: for a converging flat-space bundle the area
decreases while \(S_{\rm L}\) increases.  The expansion square is
therefore kinematic focusing; it could be read as entropy production
only under an additional coarse-graining postulate, which we state but
do not derive.

The equilibrium-Hessian sector identifies the gravitational heat with
the HWZ modified canonical-energy flux under the precise stationary,
vacuum, horizon-locking hypotheses of
Theorem~\ref{thm:secondorder}, and extends the balance to matter.
Equation \eqref{eq:Srenbalance} separates the intrinsic fixed-cell
matter source, radiative shear, kinematic focusing, and reversible cut
storage; the complete physical heat Hessian is displayed separately.
Section~\ref{sec:affine} derives the transformation laws under a
shift of the affine origin (Proposition~\ref{prop:affine} and
Table~\ref{tab:affine}).  The geometric balances retain their form
under consistent reanchoring, and a nondegenerate matched weighted
balance yields the same field equation under the hypotheses stated
there.  Positivity has an anchor-dependent domain, while the HWZ
canonical-energy identification uses the specified background
Killing field and its bifurcation anchor.
Section~\ref{sec:opensystem} records the pointwise local limit and
explains schematically how boost-cut and marginal-surface terms enter
a finite open subsystem.  Because their full stability-operator
derivation is not included here, no finite-patch MOTS area theorem is
claimed in this focused paper.

We set \(c=k_{\rm B}=\hbar=1\), retain
\(G\), and use the Hessian convention
\[
 X(\epsilon)=X_0+\epsilon\delta X+
 \frac12\epsilon^2\delta^2X+o(\epsilon^2).
\]

\section{Local causal-horizon geometry}

Fix a point \(p\) and a future-directed null vector \(k^a\) at \(p\).
Choose a small spacelike codimension-two surface through \(p\) whose
orthogonal null congruence contains \(k^a\).  A portion of the
resulting null hypersurface is denoted by \(\cH\) (see Figure 1).  Let \(V\) be an
affine parameter, with
\begin{equation}
 k^a\nabla_aV=1,\qquad k^b\nabla_bk^a=0 ,
 \label{eq:affine}
\end{equation}
and let \(C_V\) be its cross-sections.  The induced metric, area
element, expansion, and shear are \(q_{AB}\), \(\dd A\), \(\theta\),
and \(\sigma_{AB}\).  We abbreviate
\(\sigma^2:=\sigma_{AB}\sigma^{AB}\) and, for perturbations,
\(\delta\sigma^2:=\delta\sigma_{AB}\delta\sigma^{AB}\).  Thus
\begin{equation}
 \frac{\dd}{\dd V}\dd A=\theta\,\dd A .
 \label{eq:areaevolution}
\end{equation}
We choose the local equilibrium cut \(C_0\) so that
\begin{equation}
 \theta|_p=0,\qquad \sigma_{AB}|_p=0 .
 \label{eq:equilibrium}
\end{equation}
The second condition can be imposed at the selected point; it is not
assumed on a finite curved patch.

For a twist-free congruence in four dimensions, the off-shell
Raychaudhuri equation is
\begin{equation}
 \frac{\dd\theta}{\dd V}
 =-\frac12\theta^2-\sigma_{AB}\sigma^{AB}-R_{ab}k^ak^b .
 \label{eq:Raychaudhuri}
\end{equation}
No field equation has been used in \eqref{eq:Raychaudhuri}.

\subsection{Approximate boost generator}

Let \(l^a\) be the second null normal, normalized by
\(k\cdot l=-1\).  Locally inertial null normal coordinates
\((U,V,x^A)\) can be based on the transverse surface.  In these
coordinates an approximate boost vector can be chosen in the form
\begin{equation}
 \xi^a=\kappa\left(V\partial_V-U\partial_U\right)^a+\zeta^a,
 \qquad
 \zeta^a=O\!\left(\kappa\frac{\rho^3}{\cR_c^2}\right),
 \label{eq:boost}
\end{equation}
where \(\rho\) is the coordinate distance from the base point and
\(\cR_c\) is a curvature radius.  Its quadratic correction can be
set to zero.  The cubic coefficients can be chosen so that, on the
central generator \(\Gamma\),
\begin{align}
 \xi^a|_\Gamma&=\kappa Vk^a,
 \label{eq:boostongamma}\\
 \nabla_{(a}\xi_{b)}|_\Gamma&=0,
 \label{eq:Killingongamma}\\
 \nabla_a\nabla_b\xi_c|_\Gamma
 &=R^d{}_{abc}\xi_d .
 \label{eq:Killingidentity}
\end{align}
In a neighborhood,
\begin{equation}
 \nabla_{(a}\xi_{b)}
 =O\!\left(\kappa\frac{\rho^2}{\cR_c^2}\right).
 \label{eq:Killingfailure}
\end{equation}
These are precisely the orders supplied by the null-normal-coordinate
construction \cite{GuedensNNC}.  They will enter only the finite-patch
remainder, not the exact entropy identity below.

\begin{figure}[t]
\centering
\begin{tikzpicture}[scale=1.05,>=stealth,line join=round]
  \def\ra{2.45}\def\rb{0.60}
  \def\yL{3.5}\def\yV{1.85}
  \draw[->,thick] (-3.6,-0.45) -- (-3.6,\yL+0.5) node[above] {$V$};
  \draw (-3.72,0) -- (-3.48,0) node[left=6pt] {$0$};
  \draw (-3.72,\yL) -- (-3.48,\yL) node[left=6pt] {$L$};
  \fill[black!9] (-0.78,0) -- (0.78,0) -- (0.86,\yL) -- (-0.86,\yL) -- cycle;
  \draw[gray!70] (-\ra,0) -- ({-\ra*1.11},\yL);
  \draw[gray!70] (\ra,0)  -- ({\ra*1.11},\yL);
  \draw[thick] (0,0) ellipse ({\ra} and {\rb});
  \draw[thick,gray!65] (0,\yV) ellipse ({\ra*1.055} and {\rb*1.055});
  \draw[thick] (0,\yL) ellipse ({\ra*1.11} and {\rb*1.11});
  \draw[thick,black!55] (0,0) ellipse (0.78 and 0.20);
  \draw[thick,black!55] (0,\yL) ellipse (0.86 and 0.22);
  \draw[black!55] (-0.78,0) -- (-0.86,\yL);
  \draw[black!55] (0.78,0) -- (0.86,\yL);
  \draw[very thick,->] (0,0) -- (0,\yL+0.40) node[above] {$k^a$};
  \node[left] at (-0.10,2.62) {$\Gamma$};
  \fill (0,0) circle (1.7pt);
  \node[above left] at (-0.02,0.02) {$p$};
  \draw[black!55,thin] (-0.70,-0.16) -- (-1.55,-0.62);
  \node[left] at (-1.52,-0.62) {$\cD_r$};
  \node[left] at ({-\ra-0.14},0) {$C_0$};
  \node[left] at ({-\ra*1.055-0.14},\yV) {$C_V$};
  \node[left] at ({-\ra*1.11-0.14},\yL) {$C_L$};
  \node[right] at (0.95,1.05) {$\cH_{L,r}$};
  \node[right,align=left] at (2.95,2.18)
    {$\dd A(V)=J\,\dd A_0$\\[-2pt]$J>0$};
\end{tikzpicture}
\caption{Local causal horizon, fixed generator bundle, and shrinking
patch.  Cuts \(C_V\) of the null hypersurface \(\cH\) are shown at
\(V=0\), at an intermediate value, and at \(V=L\); the generators run
along \(k^a\).  The transverse widening is schematic and does not
assume \(\theta\geq0\) or monotonicity of area.  The equilibrium cut
\(C_0\), chosen so that \(\theta|_p=\sigma_{AB}|_p=0\), fixes both the
affine origin and the Lagrangian generator labels with reference
measure \(\dd A_0\).  Transport to \(C_V\) defines
\(\dd A(V)=J\,\dd A_0\), with \(J>0\) before the first caustic.
Shaded is the fixed bundle \(\cH_{L,r}\) of \eqref{eq:patch} generated
from \(\cD_r\subset C_0\); \(\Gamma\) is the central generator on which
the approximate boost \eqref{eq:boost} is exact.}
\label{fig:patch}
\end{figure}
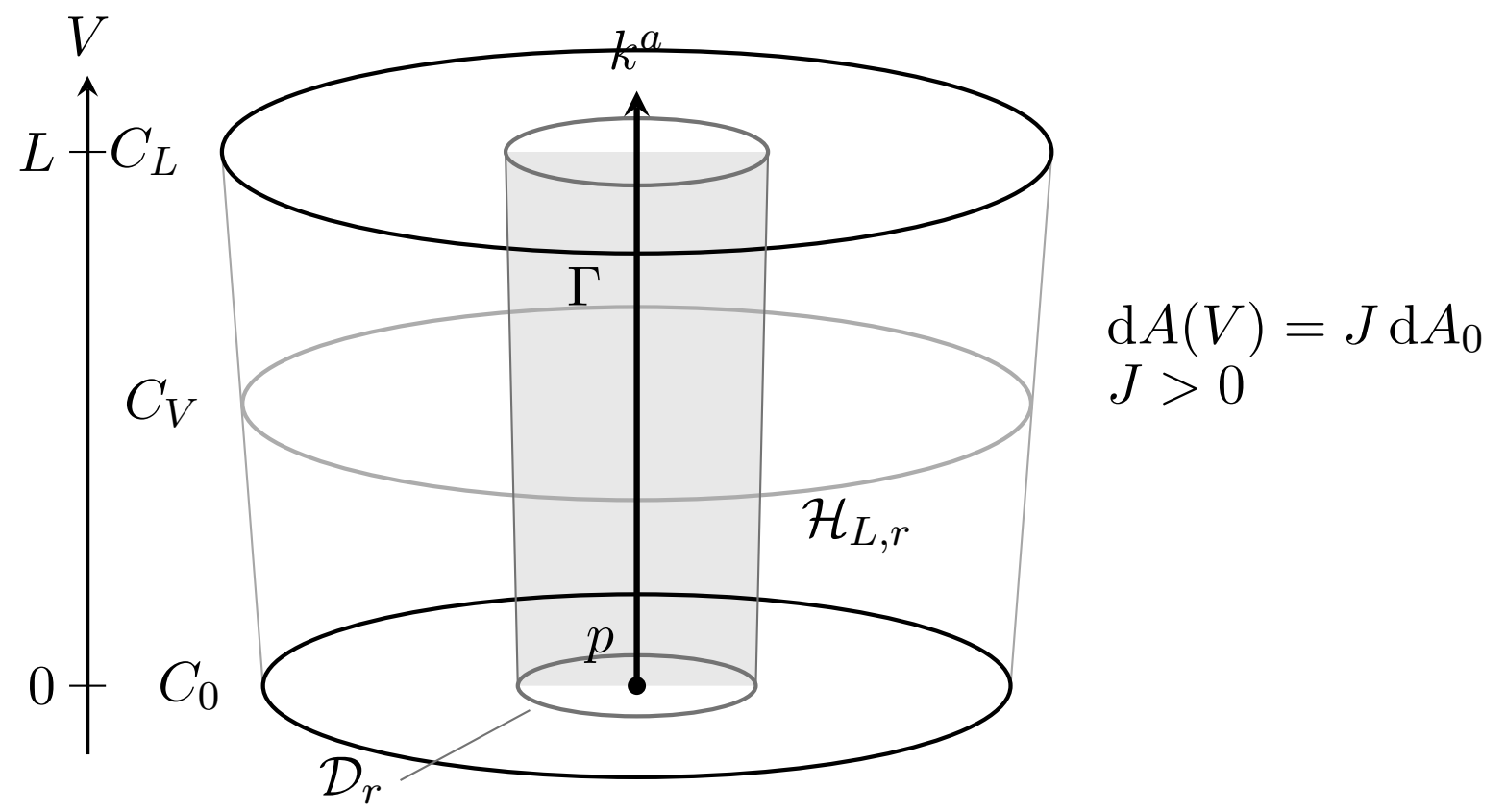

\section{An off-shell dynamical-entropy identity}
\label{sec:offshell}

On a cut \(C_V\), define the local dynamical entropy
\begin{equation}
 S_{\loc}[C_V]
 :=\frac{1}{4G}\int_{C_V}(1-V\theta)\,\dd A .
 \label{eq:Sloc}
\end{equation}
This is the Einstein-gravity HWZ expression
\(S_{\dyn}=(1-V\partial_V)S_{\rm BH}\), applied to a local causal
horizon \cite{HWZ,VisserYan}.  One point of departure should be noted
at the outset.  On a black-hole horizon the affine parameter in the
HWZ entropy is normalized so that \(V=0\) is the bifurcation surface
\cite{HWZ}.  A local causal horizon has no bifurcation surface, and
here the origin of \(V\) is instead supplied by the equilibrium cut
\(C_0\) of \eqref{eq:equilibrium}.  The entropy \eqref{eq:Sloc} is
therefore tied to a choice of equilibrium cut, which anchors the
approximate boost used in the local construction.  This is reference data rather than a
gauge redundancy.  Section~\ref{sec:affine} shows
explicitly how \(S_{\loc}\), the matter heat, and the balance laws
transform under a shift of the affine origin, which quantities are
invariant, and which change by a cut-local anchoring polarization.  Here
\eqref{eq:Sloc} is taken as a local entropy postulate; the following
identity is purely kinematic.

The form of the postulate is, however, essentially forced.  Consider
cut-local entropies of the form
\(S=(4G)^{-1}\int_{C_V}(1+f(V)\theta)\,\dd A\), reducing to the area
entropy on stationary cuts.  Differentiating along the generators and
using \eqref{eq:areaevolution},
\[
 \frac{\dd S}{\dd V}
 =\frac{1}{4G}\int_{C_V}
 \left[f\frac{\dd\theta}{\dd V}+(f'+1)\theta+f\theta^2\right]\dd A .
\]
The term linear in the instantaneous expansion vanishes for every
congruence if and only if \(f'=-1\), i.e.\ \(f=-(V-V_0)\).  Thus,
within this class, cut-locality of the first law selects the HWZ
density \eqref{eq:Sloc} uniquely, up to the choice of affine origin
\(V_0\) discussed above.  In Einstein gravity the Wall entropy
reduces to the area entropy and therefore lacks the
\(-V\theta\) correction responsible for this cancellation
\cite{HWZ}.

\begin{theorem}[Off-shell local entropy identity]
\label{thm:offshell}
For every smooth twist-free null hypersurface and fixed bundle of its
generators,
\begin{equation}
  \frac{\dd S_{\loc}}{\dd V}
 =\frac{V}{4G}\int_{C_V}
 \left(
 R_{ab}k^ak^b+\sigma_{AB}\sigma^{AB}
 -\frac12\theta^2
 \right)\dd A .
 \label{eq:offshell}
\end{equation}
\end{theorem}

\begin{proof}
Differentiate \eqref{eq:Sloc}, use
\eqref{eq:areaevolution}, and note the cancellation of the terms
linear in \(\theta\):
\begin{align}
 \frac{\dd S_{\loc}}{\dd V}
 &=-\frac{V}{4G}\int_{C_V}
 \left(\frac{\dd\theta}{\dd V}+\theta^2\right)\dd A .
 \label{eq:dSintermediate}
\end{align}
Substitution of \eqref{eq:Raychaudhuri} gives
\eqref{eq:offshell}.
\end{proof}

\begin{remark}
For the area entropy alone, the expansion must be integrated
teleologically or endpoint data must be supplied.  The
\(-V\theta\) term makes \eqref{eq:offshell} local in the cut and
removes the unwanted term linear in the instantaneous expansion.
This is the central advantage of using the dynamical entropy in a
local equation of state.
\end{remark}

\section{The first-order equation of state}

The matter boost heat through a horizon segment is
\begin{equation}
 \Delta Q_{\matt}
 :=\int_{\cH}\!T_{ab}\xi^a k^b\,\dd V\,\dd A .
 \label{eq:matterheat}
\end{equation}
On the central generator, \(\xi^a=\kappa Vk^a\), while the
temperature measured by the accelerated orbit of \(\xi^a\) is
\begin{equation}
 T_{\rm U}=\frac{\kappa}{2\pi}.
 \label{eq:Unruh}
\end{equation}
Throughout we take \(\kappa>0\); for \(\kappa=0\) the Unruh temperature
and every balance below degenerate.
Assigning this temperature to the orbits of an approximate boost on a
finite patch is an input of the construction, justified in the
shrinking-patch limit rather than derived on the patch itself;
Eq.~\eqref{eq:pointwiseshrinkinglimit} states the pointwise limit,
while \eqref{eq:Killingfailure} gives the local geometric boost error.
Neither establishes the thermality assumption on a finite patch.
The leading local heat divided by temperature is therefore
\begin{equation}
 \frac{\Delta Q_{\matt}}{T_{\rm U}}
 =2\pi\int_{\cH} V T_{ab}k^ak^b\,\dd V\,\dd A
 +\text{finite-patch remainder}.
 \label{eq:QoverT}
\end{equation}

\begin{theorem}[Conditional local equation of state]
\label{thm:Einstein}
Assume:
\begin{enumerate}
 \item every point \(p\) and null direction \(k^a\) admits local
       equilibrium patches satisfying \eqref{eq:equilibrium};
 \item the entropy density is \(1/(4G)\) and the local entropy is
       \eqref{eq:Sloc}, with affine origin at the equilibrium cut
       \(C_0\), so that the approximate boost \eqref{eq:boost}
       vanishes at \(p\), and on \(C_0\) up to the correction
       \(\zeta^a=O(\kappa\rho^3/\cR_c^2)\);
 \item the orbits of the approximate boost are assigned the Unruh
       temperature \eqref{eq:Unruh};
 \item the leading shrinking-patch balance
       \(\Delta S_{\loc}=\Delta Q_{\matt}/T_{\rm U}\) holds;
 \item \(\nabla^aT_{ab}=0\).
\end{enumerate}
Then
\begin{equation}
 G_{ab}+\Lambda g_{ab}=8\pi G T_{ab}
 \label{eq:Einstein}
\end{equation}
for a spacetime constant \(\Lambda\).
\end{theorem}

\begin{proof}
At equilibrium, the optical-square terms in \eqref{eq:offshell}
are subleading in the affine size of the patch.  Comparing
\eqref{eq:offshell} with \eqref{eq:QoverT} gives
\begin{equation}
 \left(R_{ab}-8\pi GT_{ab}\right)k^ak^b=0
 \label{eq:nullEinstein}
\end{equation}
at \(p\).  The construction applies to every null vector at every
point.  A symmetric tensor whose contraction with every null vector
vanishes is proportional to the metric, so
\begin{equation}
 R_{ab}-8\pi GT_{ab}=\Phi g_{ab}.
 \label{eq:Phi}
\end{equation}
Taking a divergence, using stress-tensor conservation and
\(\nabla^aR_{ab}=\frac12\nabla_bR\), shows that
\(\nabla_b(\Phi-\frac12R)=0\).  Hence
\(\Phi=\frac12R-\Lambda\), which is \eqref{eq:Einstein}.
\end{proof}

\begin{remark}[Where the gravitational dynamics entered]
The proof used Raychaudhuri's equation, the entropy postulate, the
matter stress tensor, the Unruh temperature, and conservation.  It
did not use the Einstein-Hilbert symplectic current, the HWZ
canonical flux, or the Einstein equation.  This is the precise
non-circularity statement.  The entropy density \(1/(4G)\) is an input
rather than an output: carrying an undetermined density \(\eta\)
through the same argument gives \(R_{kk}=(2\pi/\eta)T_{kk}\),
and hence \(G=1/(4\eta)\).  This identifies the coupling in terms
of the assumed density; it predicts neither quantity independently.
The Unruh temperature, leading Clausius relation, stress-tensor
conservation, and local equilibrium are likewise hypotheses.
The theorem is a conditional derivation of the Einstein equation
from these inputs, not a derivation of the inputs themselves.
Proposition~\ref{prop:affineEinstein} separately describes the
field-equation implication of a matched weighted balance after
reanchoring; it does not establish a physical Unruh-temperature
interpretation for an arbitrary shifted flux.
\end{remark}

\section{Exact nonlinear generator-relative balance}
\label{sec:nonlinear}

The equilibrium derivation has an exact finite-amplitude geometric
completion.  In this section the reference cut is \emph{not} assumed
to be an equilibrium cut.  We denote it by \(C_{\rm ref}\), reset the
affine origin so that \(V=0\) there, and use it to supply Lagrangian
generator labels.  Thus \(C_{\rm ref}=C_0\) only when the nonlinear
construction is expanded about the equilibrium cut of
\eqref{eq:equilibrium}.  Define the positive area Jacobian \(J\) by
\begin{equation}
 \dd A(V,x)=J(V,x)\dd A_{\rm ref}(x),
 \qquad
 J(0,x)=1,
 \qquad
 \partial_V\ln J=\theta .
 \label{eq:areaJacobian}
\end{equation}
The reference measure \(\dd A_{\rm ref}\) counts the same generator
cells at every value of \(V\).

\begin{definition}[Generator-relative functional]
\label{def:generatorrelativeentropy}
On the fixed bundle, define
\begin{equation}
  S_{\rm L}[C_V;C_{\rm ref}]
 :=
 \frac1{4G}\int_{C_{\rm ref}}
 \left(1+\ln J-V\theta\right)\dd A_{\rm ref} .
 \label{eq:generatorrelativeentropy}
\end{equation}
The subscript L denotes the Lagrangian, fixed-generator description.
The logarithm is kinematic and is unrelated to quantum logarithmic
corrections to black-hole entropy.  Following the terminology fixed in
the Introduction, \(S_{\rm L}\) is called an entropy only through its
equilibrium variations or under the postulate of
Sec.~\ref{sec:physicalentropy}.
\end{definition}

\begin{theorem}[Exact nonlinear null-sheet identity]
\label{thm:nonlinearidentity}
For every smooth twist-free, affinely parametrized null hypersurface
and every fixed generator bundle on which \(J>0\),
\begin{equation}
  \frac{\dd S_{\rm L}}{\dd V}
 =
 \frac{V}{4G}\int_{C_{\rm ref}}
 \left(
 R_{kk}+\sigma^2+\frac12\theta^2
 \right)\dd A_{\rm ref} .
 \label{eq:nonlinearoffshell}
\end{equation}
The identity is exact, local along each generator, and off shell.
\end{theorem}

\begin{proof}
At fixed generator labels,
\[
 \frac{\dd}{\dd V}(1+\ln J-V\theta)
 =-V\frac{\dd\theta}{\dd V}.
\]
The twist-free Raychaudhuri equation gives
\(-\dd\theta/\dd V=R_{kk}+\sigma^2+\theta^2/2\), proving
\eqref{eq:nonlinearoffshell}.
\end{proof}

\begin{corollary}[Monotone null focusing functional]
\label{cor:monotonefunctional}
If the null convergence condition \(R_{ab}k^ak^b\geq0\) holds on the
bundle, then
\begin{equation}
 \frac{\dd S_{\rm L}}{\dd V}\geq0
 \qquad(V\geq0)
 \label{eq:nonlinearsecondlaw}
\end{equation}
until the first caustic, with equality at a given \(V>0\) if and only
if \(R_{kk}\), \(\sigma_{AB}\), and \(\theta\) all vanish on \(C_V\).
At \(V=0\) the derivative vanishes identically, by the explicit factor
of \(V\) in \eqref{eq:nonlinearoffshell}.
This is a statement about null geometry alone: no field equation,
entropy interpretation, or coarse-graining enters.  On an Einstein
solution the null convergence condition follows from the null energy
condition.
\end{corollary}

The distinction between instantaneous-area flux and reference
normalization is essential.  Define the instantaneous-area
boost-weighted matter flux and the reference-measure conversion
remainder by
\begin{align}
 \dot{\cQ}_{\rm area}
 &:=
 \kappa V\int_{C_{\rm ref}}J T_{kk}\dd A_{\rm ref},
 \label{eq:nonlinearphysicalheat}\\
 \dot{\cC}_{\rm ref}
 &:=
 \kappa V\int_{C_{\rm ref}}(1-J)T_{kk}\dd A_{\rm ref}.
 \label{eq:referencemeasureconversion}
\end{align}
Their sum is the generator-normalized matter term
\begin{equation}
 \dot{\cQ}_{\rm L,matt}
 :=\dot{\cQ}_{\rm area}+\dot{\cC}_{\rm ref}
 =\kappa V\int_{C_{\rm ref}}T_{kk}\dd A_{\rm ref}.
 \label{eq:nonlinearLheat}
\end{equation}
The term \(\dot{\cC}_{\rm ref}\) is an algebraic Eulerian-to-Lagrangian
measure conversion, not an independently transported energy or work
channel, and it has no definite sign.  On an exact Killing horizon
\(\dot{\cQ}_{\rm area}\) is physical boost heat.  On a generic curved
local null sheet it is an instantaneous-area boost-weighted flux,
becoming physical boost heat only at leading order in the pointwise
local Rindler limit.  Finally define
\begin{equation}
 \dot{\cK}_{\rm L,opt}
 :=
 \frac{\kappa V}{8\pi G}\int_{C_{\rm ref}}
 \left(\sigma^2+\frac12\theta^2\right)\dd A_{\rm ref} .
 \label{eq:nonlinearLproduction}
\end{equation}
Here and below the measure in this expression is
\(\dd A_{\rm ref}\); the subscript L emphasizes fixed generator cells.
The shear square has a radiative interpretation near stationary
equilibrium.  The expansion square is kinematic optical focusing.

\begin{corollary}[On-shell generator-relative balance]
\label{cor:nonlinearsecondlaw}
On an Einstein solution,
\begin{equation}
  T_{\rm U}\frac{\dd S_{\rm L}}{\dd V}
 =
 \dot{\cQ}_{\rm area}
 +\dot{\cC}_{\rm ref}
 +\dot{\cK}_{\rm L,opt}.
 \label{eq:physicalgeneratorbalance}
\end{equation}
Equivalently,
\begin{equation}
 T_{\rm U}\frac{\dd S_{\rm L}}{\dd V}
 =\dot{\cQ}_{\rm L,matt}+\dot{\cK}_{\rm L,opt}.
 \label{eq:generatornormalizedbalance}
\end{equation}
Equation \eqref{eq:physicalgeneratorbalance}, rather than
\eqref{eq:generatornormalizedbalance} read in isolation, is the physical
bookkeeping statement.  It shows explicitly why the exact nonlinear
identity is not an ordinary second law formulated with physical matter
heat alone.
If \(V\geq0\) and the null energy condition holds, the null
convergence condition holds on shell and
Corollary~\ref{cor:monotonefunctional} gives monotonicity
\eqref{eq:nonlinearsecondlaw} until the first caustic.
\end{corollary}

\begin{proof}
Multiply \eqref{eq:nonlinearoffshell} by
\(T_{\rm U}=\kappa/(2\pi)\) and use the null projection of the Einstein
equation.  The null energy condition makes the matter term and both
optical squares nonnegative.
\end{proof}

The balance is in fact not merely a consequence of the field equation
but equivalent to it.

\begin{theorem}[Equivalence with the Einstein equation]
\label{thm:balanceequivalence}
Fix \(\kappa>0\) and assume \(\nabla^aT_{ab}=0\).  Then the
generator-normalized balance \eqref{eq:generatornormalizedbalance},
with \(\dot{\cK}_{\rm L,opt}\) given by
\eqref{eq:nonlinearLproduction}, has the following property if and only
if the Einstein equation holds: for every sufficiently small twist-free
generator bundle and every null direction, there exists \(V_*>0\),
prior to that bundle's first caustic, such that the balance holds for
all \(0<V<V_*\).  Explicitly, the field equation is
\begin{equation}
 G_{ab}+\Lambda g_{ab}=8\pi GT_{ab}
 \label{eq:Einsteinfromnonlinear}
\end{equation}
for a spacetime constant \(\Lambda\).
\end{theorem}

\begin{proof}
Sufficiency is Corollary~\ref{cor:nonlinearsecondlaw}.  For necessity,
subtract \eqref{eq:nonlinearLproduction} from the assumed balance and
use Theorem~\ref{thm:nonlinearidentity} to obtain
\[
 \frac{\kappa V}{8\pi G}\int_{C_{\rm ref}}
 \left(R_{kk}-8\pi GT_{kk}\right)\dd A_{\rm ref}=0 .
\]
Since \(\kappa>0\) and \(0<V<V_*\), the prefactor may be divided out.
(At \(V=0\) the balance reduces to \(0=0\) and carries no
field-equation information; this is why the hypothesis is imposed on an
interval of positive \(V\), rather than only at an unspecified
bundle-dependent isolated value of \(V\).  A single fixed common
\(V>0\) would also suffice.)  Given a point
\(q\) and a null direction there, choose the bundle and its reference
cut so that \(q\) lies at some \(V\in(0,V_*)\) on the central
generator; shrinking the bundle about that generator then gives
\(R_{ab}k^ak^b=8\pi GT_{ab}k^ak^b\) at \(q\).  Since \(q\) and the null
direction are arbitrary, this holds at every point and for every null
\(k^a\).  The null-vector argument and stress-tensor conservation then
complete the proof exactly as in Theorem~\ref{thm:Einstein}.
\end{proof}

Theorem~\ref{thm:balanceequivalence} assumes no energy condition and no
equilibrium or final cut; the null convergence condition enters only in the
monotonicity statement \eqref{eq:nonlinearsecondlaw}.  The exact
nonlinear identity of Theorem~\ref{thm:nonlinearidentity} is therefore
a finite-amplitude geometric recasting of the null-projected field
equation, normalized thermodynamically by \(T_{\rm U}\), and
characterizes that equation rather than presupposing it.  The
equivalence concerns the field equation only; it confers no entropy
status on \(S_{\rm L}\) beyond the equilibrium regime.

\subsection{Physics of the measure conversion}

The conversion term acquires a transparent interpretation upon
variation.  Consider a vacuum background matter flux,
\(T_{kk}^{(0)}=0\), and write on a fixed reference cell
\begin{equation}
 J(\epsilon)=1+\epsilon a+\frac12\epsilon^2b+o(\epsilon^2),
 \qquad
 T_{kk}(\epsilon)
 =\epsilon\tau+\frac12\epsilon^2\upsilon+o(\epsilon^2).
 \label{eq:fluxexpansions}
\end{equation}
Here \(a=\delta J=\delta\ln J\) is the first-order fractional area
strain, \(\tau=\delta T_{kk}\), and
\(\upsilon=\delta^2T_{kk}\).

\begin{proposition}[Deforming-cell flux decomposition]
\label{prop:deformingcellflux}
At fixed \(V\) and fixed reference labels,
\begin{align}
 \delta^2\dot{\cQ}_{\rm area}
 &=
 \underbrace{\kappa V\int_{C_{\rm ref}}
 \delta^2T_{kk}\dd A_{\rm ref}}_
 {\dot{\cQ}_{\rm src}^{(2)}}
 +
 \underbrace{2\kappa V\int_{C_{\rm ref}}
 a\,\delta T_{kk}\dd A_{\rm ref}}_
 {\dot{\cQ}_{\rm dil}^{(2)}},
 \label{eq:physicalheatHessiansplit}\\
 \delta^2\dot{\cC}_{\rm ref}
 &=-\dot{\cQ}_{\rm dil}^{(2)},
 \qquad
 \delta^2\dot{\cQ}_{\rm L,matt}
 =\dot{\cQ}_{\rm src}^{(2)}.
 \label{eq:conversionHessian}
\end{align}
\end{proposition}

\begin{proof}
Insert \eqref{eq:fluxexpansions} into
\eqref{eq:nonlinearphysicalheat} and
\eqref{eq:referencemeasureconversion}, and read off twice the
coefficient of \(\epsilon^2\).
\end{proof}

Thus the mixed term \(2a\,\delta T_{kk}\) is the
\emph{dilation heat}: it is the change of integrated physical flux
caused by changing the collecting area of a cell that already carries
a first-order flux.  This is the null-surface analogue of converting
transport through a deforming Eulerian control surface to transport
per fixed Lagrangian material cell.  No matter energy is created or
lost.  Because the mixed term is sign-indefinite and changes sign when
the geometric strain \(a\) is reversed at fixed matter perturbation,
it is reversible geometry--flux coupling rather than entropy
production.  The reference description subtracts the area-strain
contribution and retains the intrinsic source Hessian
\(\delta^2T_{kk}\) per background cell.  When
\(\delta T_{kk}=0\), as for a differentiable two-sided null-energy
family about vacuum, the dilation heat vanishes and the physical and
reference-cell heat Hessians coincide.

\subsection{Convex storage and the HWZ equilibrium limit}

For the affine origin selected by \(C_{\rm ref}\), define
\[
 S_{\loc}^{({\rm ref})}[C_V]
 :=\frac1{4G}\int_{C_V}(1-V\theta)\dd A.
\]
Its relation to the generator-relative functional is controlled by
\begin{equation}
 \Phi(J):=J-1-\ln J\geq0.
 \label{eq:convexareapotential}
\end{equation}
Using \(\partial_VJ=\theta J\) gives the exact storage identity
\begin{equation}
  T_{\rm U}(S_{\loc}^{({\rm ref})}-S_{\rm L})
 =
 \cB_{\rm exp}^{\rm NL},
 \qquad
 \cB_{\rm exp}^{\rm NL}
 :=
 \frac{\kappa}{8\pi G}\int_{C_{\rm ref}}
 [\Phi-V\partial_V\Phi]\dd A_{\rm ref} .
 \label{eq:nonlinearexpansionstorage}
\end{equation}
For \(J(\epsilon)=1+\epsilon a+O(\epsilon^2)\) and
\(\theta(\epsilon)=\epsilon y+O(\epsilon^2)\),
\begin{equation}
 \delta(S_{\loc}^{({\rm ref})}-S_{\rm L})=0,
 \qquad
 \delta^2\cB_{\rm exp}^{\rm NL}
 =
 \frac{\kappa}{8\pi G}\int_{C_{\rm ref}}
 (a^2-2Vay)\dd A_{\rm ref}
 =\cB_{\rm exp}.
 \label{eq:nonlinearstorageHessian}
\end{equation}
When \(C_{\rm ref}=C_0\) is the stationary equilibrium cut, the first
variation of \(S_{\rm L}\) agrees with that of the HWZ entropy, while
\begin{equation}
 \delta^2S_{\rm L}=\delta^2S_{\rm ren}.
 \label{eq:SLrenHessian}
\end{equation}
The canonical-energy law below is therefore the equilibrium Hessian
of the exact nonlinear balance.

\subsection{Finite-amplitude focusing example}

Consider a shear-free vacuum bundle with expansion constant over each
cut and reference value \(\theta_{\rm ref}\).  This is a general
reference-cut example, not an evolution from the equilibrium cut of
\eqref{eq:equilibrium}.  Raychaudhuri's equation and
\eqref{eq:areaJacobian} give
\begin{equation}
 \theta(V)=\frac{\theta_{\rm ref}}{1+\theta_{\rm ref}V/2},
 \qquad
 J(V)=\left(1+\frac{\theta_{\rm ref}V}{2}\right)^2 .
 \label{eq:homogeneousfocusing}
\end{equation}
Writing \(A_{\rm ref}=\int_{C_{\rm ref}}\dd A_{\rm ref}\), the
ordinary area and the
generator-relative functional are
\begin{align}
 A(V)
 &=A_{\rm ref}\left(1+\frac{\theta_{\rm ref}V}{2}\right)^2,
 \label{eq:homogeneousarea}\\
 S_{\rm L}(V)
 &=\frac{A_{\rm ref}}{4G}
 \left[
 1+2\ln\left(1+\frac{\theta_{\rm ref}V}{2}\right)
 -\frac{\theta_{\rm ref}V}{1+\theta_{\rm ref}V/2}
 \right].
 \label{eq:homogeneousentropy}
\end{align}
Direct differentiation gives
\begin{equation}
  \frac{\dd S_{\rm L}}{\dd V}
 =
 \frac{A_{\rm ref}V\theta_{\rm ref}^2}
 {8G(1+\theta_{\rm ref}V/2)^2}\geq0 .
 \label{eq:homogeneoussecondlaw}
\end{equation}
For \(\theta_{\rm ref}<0\), the area decreases and a caustic occurs at
\(V_{\rm c}=-2/\theta_{\rm ref}\), but \(S_{\rm L}\) increases
throughout \(0<V<V_{\rm c}\).  This elementary vacuum example proves
monotonicity of the functional but, by itself, proves neither
irreversibility nor a microscopic entropy law: the positive
\(\theta^2/2\) term is kinematic optical focusing, not gravitational
radiation.

\subsection{Structural role of the reference measure}

Consider endpoint subtractions depending only on the area Jacobian,
\[
 S_\Psi
 :=
 S_{\loc}^{({\rm ref})}
 -\frac1{4G}\int_{C_{\rm ref}}
 [\Psi(J)-V\partial_V\Psi(J)]\dd A_{\rm ref} .
\]
Writing \(X=R_{kk}+\sigma^2-\theta^2/2\), differentiation gives
\begin{equation}
 \frac{\dd S_\Psi}{\dd V}
 =
 \frac{V}{4G}\int_{C_{\rm ref}}
 [J(1-\Psi')X+J^2\Psi''\theta^2]\dd A_{\rm ref} .
 \label{eq:generalnonlinearidentity}
\end{equation}
Keeping the instantaneous-area coefficient \(J\) in front of
\(R_{kk}\) forces \(\Psi'=0\), which also removes the compensating
positive expansion term.  The choice \(\Psi=\Phi\) instead converts
both coefficients to unity and gives \eqref{eq:nonlinearoffshell}.
Thus fixed-cell normalization is structural within this
area-Jacobian endpoint class.

The endpoint computation and the linear-\(\theta\) selection of
Sec.~\ref{sec:offshell} merge into a single characterization.

\begin{theorem}[Selection of the generator-relative density]
\label{thm:selection}
Within the class
\[
 S=\frac1{4G}\int_{C_{\rm ref}}
 \bigl[1+\Psi(J)+f(V)\theta\bigr]\dd A_{\rm ref},
 \qquad
 \Psi(1)=f(0)=0,
\]
with \(\Psi\) and \(f\) smooth, suppose that (i) \(\dd S/\dd V\)
contains no term linear in the instantaneous expansion, for every
twist-free congruence, and (ii) its integrand is pointwise nonnegative
for \(V\geq0\) under the null convergence condition.  Then
\[
 \Psi(J)=\alpha\ln J,\qquad f(V)=-\alpha V,\qquad \alpha\geq0 .
\]
First-order agreement with the HWZ dynamical entropy about an
equilibrium reference cut fixes \(\alpha=1\), i.e.\ \(S=S_{\rm L}\)
uniquely.
\end{theorem}

\begin{proof}
At fixed generator labels, \(\partial_VJ=\theta J\) and Raychaudhuri's
equation give
\[
 \frac{\dd S}{\dd V}
 =\frac1{4G}\int_{C_{\rm ref}}
 \Bigl[\bigl(J\Psi'(J)+f'(V)\bigr)\theta
 -f(V)\Bigl(\tfrac12\theta^2+\sigma^2+R_{kk}\Bigr)\Bigr]
 \dd A_{\rm ref}.
\]
Off shell, \(R_{kk}\) is unrestricted, so \(J\) and \(\theta\) may be
prescribed independently on a cut: given targets \(J_*\) and
\(\theta_*\) at \(V\), choose a smooth history \(\theta(V')\) on
\([0,V]\) satisfying both \(\theta(V)=\theta_*\) and
\(\int_0^V\theta(V')\,\dd V'=\ln J_*\), set \(\sigma_{AB}=0\), and
\emph{determine} \(R_{kk}\) along the generator from Raychaudhuri's
equation.  Throughout, the
reference labels and the measure \(\dd A_{\rm ref}\) are held fixed.  Condition (i) therefore forces
\(J\Psi'(J)+f'(V)=0\) for all \(J\) and \(V\) separately, whence
\(J\Psi'(J)=\alpha\) and \(f'(V)=-\alpha\) for a constant \(\alpha\); the
normalizations \(\Psi(1)=f(0)=0\) give \(\Psi=\alpha\ln J\) and
\(f=-\alpha V\).  Condition (ii) applied to the remaining squares requires
\(\alpha\geq0\).  Finally, about an equilibrium reference cut the first
variation is \(\delta S=\alpha\,\delta S_{\rm L}\), and
\(\delta S_{\rm L}=\delta S_{\rm HWZ}\) by
\eqref{eq:entropicanchor}, so agreement with the HWZ physical-process
first law fixes \(\alpha=1\).
\end{proof}

Condition (i) is the same cut-locality requirement that selected the
\(-V\theta\) correction in Sec.~\ref{sec:offshell}; the theorem shows
that, on fixed generator cells, it simultaneously selects the
\(\ln J\) storage term.  The normalization \(f(0)=0\) places the
affine origin on the reference cut, and it can be dropped.  With \(f(0)\) free, the general solution is
\(\Psi=\alpha\ln J\), \(f=-\alpha V+\beta\).  Condition (ii) on
\(V\geq0\) requires \(\alpha\geq0\) and \(\beta\leq0\);
the branch \(\alpha=0,\ \beta<0\) is allowed by these geometric
conditions.  First-order agreement with the HWZ representative for
general admissible variations fixes \(\alpha=1\) and \(\beta=0\).
Indeed, the variation is
\(\alpha\,\delta S_{\rm L}+(\beta/4G)\int\delta\theta\,\dd A_{\rm ref}\);
agreement is imposed for all variations, including independent
area-strain and expansion responses on the evaluation cut.
For \(\alpha>0\) one may write \(f=-\alpha(V-V_0)\) with
\(V_0=\beta/\alpha\leq0\); HWZ agreement then gives \(V_0=0\).
Agreement restricted only to the vacuum sector with
\(\delta\theta=0\) would not determine this shift.
With the stated general matching requirement, the theorem
therefore also fixes the affine origin at the reference cut
(Sec.~\ref{sec:affine}).\footnote{Condition (i) cannot be replaced by
on-shell monotonicity alone.  Under the null convergence condition the
expansion is nonincreasing along each generator, so the realizable
pairs \((J,\theta)\) on a cut satisfy \(J\geq e^{V\theta}\); this
constraint leaves the linear-\(\theta\) coefficient undetermined on
part of the \(J<1\) region.  Imposing (i) off shell, where \(R_{kk}\)
is unrestricted and \(J\) and \(\theta\) are independent, removes this
slack.}  The result remains restricted to the stated
two-function class and does not establish uniqueness among all
possible horizon entropies.

\subsection{Physical entropy status}
\label{sec:physicalentropy}

There is a physically anchored perturbative statement and a
conditional finite-amplitude extension.  For
\(C_{\rm ref}=C_0\) and a perturbation about stationary equilibrium,
\begin{equation}
 \delta S_{\rm L}=\delta S_{\loc}=\delta S_{\rm HWZ},
 \qquad
 \delta^2S_{\rm L}=\delta^2S_{\rm ren}.
 \label{eq:entropicanchor}
\end{equation}
The first equality ties the construction to the HWZ physical-process
first law.  In semiclassical gravity, HWZ relate their dynamical
gravitational entropy, together with a modified matter von Neumann
entropy, to the generalized second law and the quantum null energy
condition \cite{HWZ,QNEC}.  Replica calculations further identify the
linearized HWZ entropy with gravitational entropy across the
perturbative apparent horizon \cite{JiaQiGao}.  The second equality in
\eqref{eq:entropicanchor}, completed by
Theorem~\ref{thm:secondorder}, identifies the temperature-weighted change of the vacuum equilibrium
Hessian with the positive gravitational canonical-energy flux.  Consequently,
the entropy interpretation of \(S_{\rm L}\) is physically anchored
through its first variation and its renormalized quadratic response
about equilibrium.  This perturbative agreement is the strongest entropy statement
established here, under the hypotheses of that theorem.

A finite-amplitude interpretation requires additional coarse-graining.
We consider an observer restricted to one side of the local causal
horizon, using the Unruh state as the local equilibrium reference.
This causal restriction is motivated by generalized-second-law
results for Rindler horizons \cite{WallGSL}, but those results do not
derive the functional below.
The observer retains the reference generator cells and the
instantaneous macroscopic fields \(J\) and \(\theta\), but traces over
matter and gravitational correlations across the horizon and discards
the optical history and transverse shape data that lead to the same
macroscopic pair \((J,\theta)\).  Evolution is stopped at the first
caustic, where the fixed-cell description ceases to be one-to-one.
This is a proposed causal, Lagrangian coarse-graining, not a
constructed map on microscopic states.  We \emph{postulate}
\(S_{\rm L}\) as a possible gravitational entropy potential for
such a description; the causal restriction alone does not derive it.
It then has the standard thermodynamic consistency properties:
additivity on disjoint generator bundles, normalization to \(A/(4G)\)
on the chosen equilibrium reference cut, invariance under generator
relabeling and affine rescaling, and compatibility with the first law
through \eqref{eq:entropicanchor}.  It is not invariant under a shift
of the affine origin; that dependence is part of the postulate's
anchoring data (Sec.~\ref{sec:affine}).  If the postulate is granted,
monotonicity \eqref{eq:nonlinearsecondlaw} would express a second law
for this coarse-grained description under the null convergence
condition.  Because the postulate is not derived here, we do not
claim an \(H\)-theorem.

The qualifier is indispensable.  If the complete optical data are
retained and the congruence is treated as a reversible classical
system, the expansion square in a shear-free vacuum bundle is not
microscopic dissipation.  In that description \(S_{\rm L}\) is a
geometric Lyapunov functional.  It could admit a physical
coarse-grained entropy interpretation if the additional postulate for
the causally restricted observer is justified.
Thus this paper physically anchors the near-equilibrium entropy
response and states a finite-amplitude entropy proposal whose
physical status remains open; it does not claim a nonperturbative
replica or state-counting derivation.

\section{The second-order gravitational completion}

We may now work on the solution space of \eqref{eq:Einstein}.  The
geometric setting also changes at this point, and the change should be
stated explicitly.  The preceding sections concern a generic curved
local causal horizon, on which no exact boost Killing field exists;
there the approximate generator \eqref{eq:boost} and the finite-patch
remainder are required.  This section and the matter Hessian below
instead use a stationary vacuum bifurcate Killing horizon carrying an
exact Killing generator with constant surface gravity.  The
approximate-boost error \eqref{eq:Killingfailure} therefore does not
enter those exact second-order results.  Section~\ref{sec:opensystem}
then returns to the generic curved local horizon and distinguishes the
pointwise shrinking limit from stronger finite-patch estimates not
claimed here.  On the stationary horizon the affine origin is
fixed by the bifurcation surface exactly as in \cite{HWZ}.
Equation~\eqref{eq:affineHessian} below shows that this choice is
used in identifying the temperature-weighted change of the second
variation with the canonical-energy flux of the fixed Killing field
\(\xi\) that defines \(e_G\).

Consider a one-parameter family
\begin{equation}
 g_{ab}(\epsilon)
 =g^{(0)}_{ab}+\epsilon h_{ab}
 +\frac12\epsilon^2j_{ab}+O(\epsilon^3)
 \label{eq:family}
\end{equation}
about a stationary vacuum bifurcate Killing horizon.  Use a
horizon-locking gauge, take \(V\) to be the background affine
parameter on its future branch, and assume a first-order vacuum
perturbation.  On the background,
\begin{equation}
 \theta^{(0)}=0,\qquad \sigma^{(0)}_{AB}=0,\qquad
 R^{(0)}_{kk}=0.
 \label{eq:Killingbackground}
\end{equation}
If \(\delta\theta=0\) on the initial cut, the linearized
Raychaudhuri equation and \(\delta R_{kk}=0\) give
\(\delta\theta=0\) along the horizon segment.

\begin{theorem}[Second-order shear balance and canonical-energy flux]
\label{thm:secondorder}
Let the family \eqref{eq:family} solve the vacuum Einstein equation
with fixed cosmological constant through second order on the
future-horizon segment \(\cH_{12}\), where
\(0\leq V_1<V_2\), and impose the conditions above.  Then
\begin{equation}
  \delta^2S_{\loc}[C_{V_2}]
 -\delta^2S_{\loc}[C_{V_1}]
 =\frac{1}{2G}
 \int_{V_1}^{V_2}\!\!\int_{C_V}
 V\,\delta\sigma_{AB}\delta\sigma^{AB}
 \,\dd A_0\,\dd V .
 \label{eq:secondentropy}
\end{equation}
Consequently,
\begin{equation}
  T_{\rm U}\,\Delta\delta^2S_{\loc}
 =\frac{\kappa}{4\pi G}
 \int_{\cH_{12}}V\,\delta\sigma^2\,\dd V\,\dd A_0
 =\int_{\cH_{12}}e_G[h,h]\geq0 .
 \label{eq:canonicalidentity}
\end{equation}
\end{theorem}

\begin{proof}
Take two variations of \eqref{eq:offshell}.  The background optical
scalars vanish and \(\delta\theta=0\).  In the horizon-locking gauge
\(k^a\) remains null, while the vacuum family obeys
\(R_{ab}(\epsilon)=\Lambda g_{ab}(\epsilon)\).  Hence
\(R_{kk}(\epsilon)=0\) identically and both
\(\delta R_{kk}\) and \(\delta^2R_{kk}\) vanish.  Therefore
\begin{equation}
 \frac{\dd}{\dd V}\delta^2S_{\loc}
 =\frac{V}{2G}\int_{C_V}
 \delta\sigma_{AB}\delta\sigma^{AB}\,\dd A_0 .
 \label{eq:dsecondentropy}
\end{equation}
Integration gives \eqref{eq:secondentropy}.  Multiplication by
\(T_{\rm U}=\kappa/(2\pi)\) gives the first equality in
\eqref{eq:canonicalidentity}.  In Einstein gravity the HWZ modified
canonical-energy flux on a Killing horizon is
\begin{equation}
 e_G[h,h]
 =\frac{1}{4\pi G}(\kappa V)^2
 \delta\sigma_{AB}\delta\sigma^{AB}\,
 \epsilon^{(3)} ,
 \qquad
 \epsilon^{(3)}
 =\frac{\dd V\,\dd A_0}{\kappa V}.
 \label{eq:eG}
\end{equation}
This proves the second equality.  Positivity is manifest.
\end{proof}

\begin{remark}
The condition \(\delta\theta=0\) on the initial cut is a choice of
equilibrium initial data, not a consequence of the field equations.
The bulk shear flux in \eqref{eq:secondentropy} is independent of this
choice; what depends on it is the absence of expansion-storage
endpoint terms, which reappear for general initial data as the
cut polarization of Secs.~\ref{sec:matter} and
\ref{sec:opensystem}.
\end{remark}

\begin{corollary}[Second-order monotonicity]
For a nonzero first-order radiative perturbation, on segments to the
future of the bifurcation cut (\(V_1\geq0\); cf.\
\eqref{eq:affineHessian}),
\begin{equation}
 \Delta\delta^2S_{\loc}\geq0,
 \label{eq:secondlaw}
\end{equation}
with equality precisely when the horizon shear perturbation vanishes
almost everywhere on the segment, modulo degenerate gauge
directions.
\end{corollary}

\section{Matter-coupled second-order balance}
\label{sec:matter}

The Einstein equation may now be used in the exact identity
\eqref{eq:offshell}.  On any Einstein solution,
\begin{equation}
  T_{\rm U}\frac{\dd S_{\loc}}{\dd V}
 =
 \kappa V\int_{C_V}T_{kk}\,\dd A
 +\frac{\kappa V}{8\pi G}\int_{C_V}
 \left(\sigma^2-\frac12\theta^2\right)\dd A .
 \label{eq:exactmatterbalance}
\end{equation}
The first term is the instantaneous-area boost-weighted matter flux.
It is physical boost heat on an exact Killing horizon and at leading
order in the pointwise local Rindler limit.  The second is the exact
nonlinear optical correction.  Unlike the off-shell identity used in
Theorem~\ref{thm:Einstein}, equation
\eqref{eq:exactmatterbalance} is an on-shell consequence of the
Einstein equation already obtained.

\subsection{Complete Hessian in fixed-boost gauge}

Take a stationary vacuum bifurcate-Killing-horizon background and a
fixed bundle of generators on its future branch.  Use a
horizon-locking gauge in which \(V\), \(k^a\),
\(\xi^a=\kappa Vk^a\), and the generator labels are fixed.  Define
\begin{equation}
 \delta(\dd A)=a\,\dd A_0,\qquad
 \frac{\dd a}{\dd V}=\delta\theta=:y .
 \label{eq:aydefinition}
\end{equation}
Since \(T_{kk}^{(0)}=0\),
\begin{equation}
 \delta^2(T_{kk}\dd A)
 =\left(\delta^2T_{kk}+2a\,\delta T_{kk}\right)\dd A_0.
 \label{eq:mattermeasurevariation}
\end{equation}
Accordingly, the complete physical boost-heat Hessian is the sum of
the intrinsic fixed-cell source and the dilation heat of
Proposition~\ref{prop:deformingcellflux}:
\begin{equation}
 \dot{\cQ}_{\rm phys}^{(2)}
 =
 \dot{\cQ}_{\rm src}^{(2)}
 +\dot{\cQ}_{\rm dil}^{(2)}
 =
 \kappa V\int_{C_V}
 \left(\delta^2T_{kk}+2a\,\delta T_{kk}\right)\dd A_0 .
 \label{eq:completephysicalheatHessian}
\end{equation}
Only \(\dot{\cQ}_{\rm phys}^{(2)}\), not
\(\dot{\cQ}_{\rm src}^{(2)}\) alone, is the complete physical heat
response when \(\delta T_{kk}\neq0\).

\begin{theorem}[Matter-coupled horizon Hessian]
\label{thm:matterhessian}
Under these conditions,
\begin{align}
 T_{\rm U}\frac{\dd}{\dd V}\delta^2S_{\loc}
 ={}&
 \kappa V\int_{C_V}
 \left(\delta^2T_{kk}+2a\,\delta T_{kk}\right)\dd A_0
 \notag\\
 &+\frac{\kappa V}{4\pi G}\int_{C_V}
 \left(\delta\sigma^2-\frac12y^2\right)\dd A_0 .
 \label{eq:matterhessian}
\end{align}
\end{theorem}

\begin{proof}
Take two variations of \eqref{eq:exactmatterbalance}.  The
background stress tensor and optical scalars vanish.  Equation
\eqref{eq:mattermeasurevariation} gives the matter terms, while
\(\delta^2(\sigma^2)=2\delta\sigma^2\) and
\(\delta^2(\theta^2)=2y^2\) give the optical terms.
\end{proof}

The four-dimensional integrand in \eqref{eq:matterhessian} has the
same structural form as the local horizon part of the matter-sensitive
modified canonical energy of Ref.~\cite{FuJiZhouCao}, with
corresponding shear, expansion, and area-weighted stress-tensor terms;
we do not attempt a term-by-term identification of conventions here.
In either treatment the factor \(2a\,\delta T_{kk}\) is required:
omitting variation of the area measure gives an incomplete matter
Hessian.  Its meaning is not a second matter species or an additional
transport channel; it is the area-strain correction to flux through
the perturbed cut.

\subsection{Expansion storage and the renormalized balance}

The linearized Raychaudhuri equation gives
\begin{equation}
 \frac{\dd y}{\dd V}=-8\pi G\,\delta T_{kk}.
 \label{eq:linearRaymatter}
\end{equation}
Consequently,
\begin{equation}
 2a\,\delta T_{kk}-\frac{1}{8\pi G}y^2
 =
 \frac{1}{8\pi G}
 \left[y^2-2\frac{\dd}{\dd V}(ay)\right].
 \label{eq:mixedreorganization}
\end{equation}
The apparently negative expansion contribution therefore cannot be
interpreted separately from the dilation part of the physical heat.

Define the reversible cut-storage functional
\begin{equation}
 \cB_{\rm exp}[C_V]
 :=
 \frac{\kappa}{8\pi G}\int_{C_V}
 \left(a^2-2Vay\right)\dd A_0 .
 \label{eq:Bexp}
\end{equation}
Integrating \eqref{eq:matterhessian}, using
\eqref{eq:mixedreorganization}, gives
\begin{equation}
 \begin{aligned}
 T_{\rm U}\Delta\delta^2S_{\loc}
 ={}&
 \kappa\int_{\cH_{12}}V\,\delta^2T_{kk}\,\dd V\,\dd A_0\\
 &+\frac{\kappa}{4\pi G}
 \int_{\cH_{12}}V\,\delta\sigma^2\,\dd V\,\dd A_0\\
 &+\frac{\kappa}{8\pi G}
 \int_{\cH_{12}}V\,(\delta\theta)^2\,\dd V\,\dd A_0\\
 &+\cB_{\rm exp}[C_{V_2}]-\cB_{\rm exp}[C_{V_1}] .
 \end{aligned}
 \label{eq:renormalizedmatterbalance}
\end{equation}
Thus the dilation heat and the negative expansion square reorganize,
on shell, into positive bulk focusing and reversible storage on the
endpoint cuts.  The dilation contribution has not disappeared: the
linearized focusing equation has converted it into a bulk optical term
and a boundary strain polarization.  Equation
\eqref{eq:renormalizedmatterbalance} is an identity, not a positivity
claim for arbitrary external sources: \(\delta^2T_{kk}\) need not
have a sign.

Equivalently, define a cut-renormalized entropy Hessian by
\begin{equation}
 T_{\rm U}\delta^2S_{\rm ren}[C_V]
 :=T_{\rm U}\delta^2S_{\loc}[C_V]-\cB_{\rm exp}[C_V].
 \label{eq:Sren}
\end{equation}
Its balance contains only bulk terms:
\begin{equation}
  T_{\rm U}\Delta\delta^2S_{\rm ren}
 =
 \int_{\cH_{12}}\!\!
 \left[
 \kappa V\delta^2T_{kk}
 +\frac{\kappa V}{4\pi G}\delta\sigma^2
 +\frac{\kappa V}{8\pi G}(\delta\theta)^2
 \right]\dd V\,\dd A_0 .
 \label{eq:Srenbalance}
\end{equation}
The first term in \eqref{eq:Srenbalance} is the intrinsic fixed-cell
source Hessian, not the complete physical heat Hessian for
\(\delta T_{kk}\neq0\).  The subtraction in \eqref{eq:Sren} is a
boundary polarization; it is not a new microscopic entropy postulate.

\subsection{Null energy condition and a scalar-field test}

Suppose a differentiable two-sided family satisfies the null energy
condition with \(T_{kk}(0)=0\).  Then \(\epsilon=0\) is a local
minimum of \(T_{kk}(\epsilon)\), and hence
\begin{equation}
 \delta T_{kk}=0,\qquad \delta^2T_{kk}\geq0.
 \label{eq:NEChessian}
\end{equation}
With equilibrium initial data, \eqref{eq:linearRaymatter} implies
\(\delta\theta=0\); fixing the initial area also gives \(a=0\).
Therefore \(\cB_{\rm exp}=0\), and
\begin{equation}
  T_{\rm U}\Delta\delta^2S_{\loc}
 =
 \int_{\cH_{12}}
 \left[
 \kappa V\delta^2T_{kk}
 +\frac{\kappa V}{4\pi G}\delta\sigma^2
 \right]\dd V\,\dd A_0
 \geq0 .
 \label{eq:mattersecondlaw}
\end{equation}
This is the local matter-plus-gravity second law for perturbations
that enter quadratically about a vacuum background.

For an explicit test, take a minimally coupled scalar
\(\phi(\epsilon)=\epsilon\varphi+
\frac12\epsilon^2\psi+O(\epsilon^3)\) about \(\phi_0=0\).  Since
\begin{equation}
 T_{kk}=(k^a\nabla_a\phi)^2,
 \qquad
 \delta T_{kk}=0,\qquad
 \delta^2T_{kk}=2(k^a\nabla_a\varphi)^2,
 \label{eq:scalarTkk}
\end{equation}
the potential drops out and \eqref{eq:mattersecondlaw} becomes
\begin{equation}
  T_{\rm U}\Delta\delta^2S_{\loc}
 =
 2\kappa\int_{\cH_{12}}V(k\varphi)^2\,\dd V\,\dd A_0
 +\frac{\kappa}{4\pi G}
 \int_{\cH_{12}}V\,\delta\sigma^2\,\dd V\,\dd A_0
 \geq0 .
 \label{eq:scalarbalance}
\end{equation}
The first term is the Hessian-normalized scalar boost-energy flux;
the second is the graviton/HWZ flux.  The physical quadratic Taylor
contribution is one half of \eqref{eq:scalarbalance}.

\section{Affine-origin dependence: invariants and anchoring polarizations}
\label{sec:affine}

The constructions above use an affine normalization, an anchoring
cut, and, for \(S_{\rm L}\), a reference-cell measure.  The latter two
are distinct data even when supplied by the same cut.  Throughout
this section, comparisons use the same geometric evaluation cuts
and generator bundle.  Rescaling and anchoring parameters are held
fixed under perturbative variations.

\paragraph{Passive relabeling and rescaling.}
With an anchor at coordinate value \(V_0\), the affine weight is
\(v=V-V_0\).  A passive relabeling \(W=V+b\) also gives
\(W_0=V_0+b\); hence \(W-W_0=v\).  This changes neither functional
nor the integrated balances.  It must be distinguished from changing
the physical anchoring while holding the evaluation cuts fixed.

Under \(k^a\to k^a/\lambda\), \(V\to\lambda V\), with constant
\(\lambda>0\), one has
\(\theta\to\theta/\lambda\),
\((\sigma^2,R_{kk},T_{kk})\to(\sigma^2,R_{kk},T_{kk})/\lambda^2\),
and \(\dd V\to\lambda\dd V\).  The geometric cuts, their area
measures, and \(J\) are unchanged when compared at the corresponding
coordinate values.  Thus \(v\theta\) and \(vR_{kk}\dd V\) are
invariant.  With \(\kappa\) fixed, the on-horizon boost weight
\(\kappa v k^a\), \(T_{\rm U}\), the functionals, and the integrated
balances are invariant; rate equations scale as \(1/\lambda\).
Their perturbative variations inherit this invariance when
\(\lambda\) is independent of the perturbation parameter.
Generator-dependent rescalings preserve affinity but change the
constant-coordinate foliation; only constant rescalings are considered
here.

\paragraph{Physical reanchoring at fixed reference measure.}
Let \(C_c\) be the cut \(V=c\), with \(c\) constant.  Replacing the
weight \(V\) by \(V-c\) changes the anchor to \(C_c\).
For the shift \(V\mapsto V+c\), replace \(c\) with \(-c\)
in the following formulas.  First retain
\(J=\dd A/\dd A_{\rm ref}\) and the old reference measure:
\begin{align}
 S_{\loc}^{(c)}[C_V]
 &:=\frac1{4G}\int_{C_V}\bigl(1-(V-c)\theta\bigr)\dd A,
 \label{eq:Slocc}\\
 S_{\rm L}^{(c)}[C_V;C_{\rm ref}]
 &:=\frac1{4G}\int_{C_{\rm ref}}
 \bigl(1+\ln J-(V-c)\theta\bigr)\dd A_{\rm ref}.
 \label{eq:SLc}
\end{align}
For the algebraic flux comparison, define on the null sheet
\begin{equation}
 \xi_c^a:=\xi^a-c\kappa k^a
          =\kappa(V-c)k^a+\zeta^a,\qquad
 \Delta Q_{\matt}^{(c)}
 :=\int_{\cH}T_{ab}\xi_c^ak^b\,\dd V\,\dd A .
 \label{eq:xic}
\end{equation}
The correction \(\zeta^a\) in this definition is the original one.
Thus \(\Delta Q_{\matt}^{(c)}\) is a weighted matter-flux
functional; its interpretation as physical heat is not established
for an arbitrary \(c\).  In particular
\(\xi_c|_{C_c}=\zeta|_{C_c}\) need not vanish across the cut,
and subtracting \(c\kappa k^a\) does not in general preserve the
approximate Killing properties.  A thermal interpretation requires
a suitable approximate-boost construction and temperature assumption.
On a planar Rindler horizon in Minkowski space, the shifted horizon
weight does extend to an exact translated boost.  This special
property is not asserted for an arbitrary null sheet, even in flat
spacetime.  When \(0<c<L\), \(V-c\) changes sign across the patch:
the formulas remain valid as signed weighted identities, but do not
describe one future-directed boost over the entire segment.

Define the expansion functionals and unweighted matter flux by
\begin{equation}
 \Pi[C_V]:=\frac1{4G}\int_{C_V}\theta\,\dd A
 =\frac{\dd S_{\rm BH}}{\dd V},\qquad
 \Pi_{\rm L}[C_V]:=\frac1{4G}\int_{C_{\rm ref}}\theta\,\dd A_{\rm ref},
 \qquad
 \Delta P:=\int_{\cH}T_{kk}\,\dd V\,\dd A .
 \label{eq:anchoringcharges}
\end{equation}
Here \(S_{\rm BH}=A/(4G)\).  The shifts \(c\Pi\) and
\(c\Pi_{\rm L}\) will be called cut-local anchoring polarizations,
in entropy units.  Their temperature-weighted endpoint differences
enter the energy-normalized balances.  This terminology identifies
their role in the identities; no separately derived Hamiltonian or
Noether translation charge, or conservation law for \(\Pi\), is
asserted.  Under rescaling, \(\Pi,\Pi_{\rm L},\Delta P\) scale as
\(1/\lambda\) and \(c\to\lambda c\), so their products with \(c\)
are invariant.

\begin{proposition}[Affine-shift transformation laws]
\label{prop:affine}
For every smooth twist-free bundle with \(J>0\) and constant \(c\):
\begin{enumerate}
 \item[(a)] The functionals change by anchoring polarizations,
 \begin{equation}
  S_{\loc}^{(c)}=S_{\loc}+c\Pi,\qquad
  S_{\rm L}^{(c)}=S_{\rm L}+c\Pi_{\rm L}.
  \label{eq:affineshiftS}
 \end{equation}
 \item[(b)] The exact identities retain their form:
 \begin{align}
  \frac{\dd S_{\loc}^{(c)}}{\dd V}
  &=\frac{V-c}{4G}\int_{C_V}
    \left(R_{kk}+\sigma^2-\frac12\theta^2\right)\dd A,
  \label{eq:offshellc}\\
  \frac{\dd S_{\rm L}^{(c)}}{\dd V}
  &=\frac{V-c}{4G}\int_{C_{\rm ref}}
    \left(R_{kk}+\sigma^2+\frac12\theta^2\right)\dd A_{\rm ref}.
  \label{eq:nonlinearc}
 \end{align}
 \item[(c)] For the flux defined in \eqref{eq:xic},
 \begin{equation}
  \frac{\Delta Q_{\matt}^{(c)}}{T_{\rm U}}
  =\frac{\Delta Q_{\matt}}{T_{\rm U}}-2\pi c\,\Delta P .
  \label{eq:affineshiftQ}
 \end{equation}
 \item[(d)] The formal Clausius defect
 \(\mathfrak D^{(c)}
 :=\Delta S_{\loc}^{(c)}-\Delta Q_{\matt}^{(c)}/T_{\rm U}\)
 satisfies
 \begin{align}
  \mathfrak D^{(c)}
  &=\mathfrak D^{(0)}+c\bigl(\Delta\Pi+2\pi\Delta P\bigr),
  \notag\\
  \Delta\Pi+2\pi\Delta P
  &=-\frac1{4G}\int_{\cH}
  \left[(R_{kk}-8\pi GT_{kk})+\sigma^2-\frac12\theta^2\right]
  \dd V\,\dd A .
  \label{eq:affineshiftdefect}
 \end{align}
\end{enumerate}
\end{proposition}

\begin{proof}
Part (a) follows from the definitions.  For (b), differentiate the
densities using \(\partial_V\ln J=\theta\),
\(\partial_V\dd A=\theta\dd A\), and Raychaudhuri's equation.
Part (c) follows from \(\xi_c-\xi=-c\kappa k\) and
\(T_{\rm U}=\kappa/(2\pi)\).  For (d), combine (a) and (c) with
\(\partial_V\Pi=(4G)^{-1}\int(\partial_V\theta+\theta^2)\dd A\).
\end{proof}

The purely geometric density formulas in (a) and (b) also hold for
\(c=c(x^A)\) constant along each generator, provided \(c\) stays
inside the integrals.  No approximate-boost interpretation of such a
generator-dependent reanchoring is assumed here.

On an Einstein solution, the cut-local balance becomes
\begin{equation}
 T_{\rm U}\frac{\dd S_{\loc}^{(c)}}{\dd V}
 =\kappa(V-c)\int_{C_V}T_{kk}\dd A
 +\frac{\kappa(V-c)}{8\pi G}\int_{C_V}
       \left(\sigma^2-\frac12\theta^2\right)\dd A .
 \label{eq:exactmatterbalancec}
\end{equation}
The generator-relative balance likewise follows from (b), with the
reference measure and the positive expansion-square term.  Equations
\eqref{eq:nonlinearphysicalheat}--\eqref{eq:nonlinearLproduction}
transform by replacing their weight \(V\) with \(V-c\); their
sum changes by \(T_{\rm U}c\,\partial_V\Pi_{\rm L}\).
Between fixed endpoints the resulting change is
\(T_{\rm U}c\,\Delta\Pi_{\rm L}\).  For the cut-local balance the
corresponding change is
\begin{equation}
 T_{\rm U}c\,\Delta\Pi
 =-c\kappa\Delta P
  -\frac{c\kappa}{8\pi G}\int_{\cH}
       \left(\sigma^2-\frac12\theta^2\right)\dd V\,\dd A .
 \label{eq:polarizationbudget}
\end{equation}
Thus \(-c\kappa\Delta P\) is the matter part of the change; the
optical contribution is also required.  The individual terms depend
on the anchor even though the balance retains its form.
For \(\Phi=J-1-\ln J\), the nonlinear storage shifts as
\begin{equation}
 \cB_{\rm exp}^{{\rm NL},(c)}-\cB_{\rm exp}^{\rm NL}
 =\frac{\kappa c}{8\pi G}
       \int_{C_{\rm ref}}\partial_V\Phi\,\dd A_{\rm ref}
 =T_{\rm U}c(\Pi-\Pi_{\rm L}).
 \label{eq:affinestorage}
\end{equation}

\paragraph{Changing the reference cut as well.}
Moving the reference measure to \(C_c\) replaces \(J\) by \(J/J_c\)
and \(\dd A_{\rm ref}\) by \(\dd A_c=J_c\dd A_{\rm ref}\),
where \(J_c:=J(c,x)>0\).  Consequently
\begin{equation}
 S_{\rm L}[C_V;C_c]-S_{\rm L}[C_V;C_{\rm ref}]
 =\frac1{4G}\int_{C_{\rm ref}}
 \left[(J_c-1)\bigl(1+\ln J-(V-c)\theta\bigr)
       -J_c\ln J_c+c\theta\right]\dd A_{\rm ref}.
 \label{eq:refchange}
\end{equation}
Its derivative is
\((4G)^{-1}\int_{C_c}(V-c)
(R_{kk}+\sigma^2+\theta^2/2)\dd A_c\).
The corresponding on-shell balance uses the same new measure.
This is a change of cell weights as well as of affine origin, so
the functional changes by more than \(c\Pi_{\rm L}\).

\paragraph{Field-equation content of a matched weighted balance.}
The following statement concerns the weighted source functional
\eqref{eq:xic}.  It does not establish Unruh thermality for a new
observer at each anchor.

\begin{proposition}[Nondegenerate weighted balance and the field equation]
\label{prop:affineEinstein}
Keep the equilibrium conditions at \(p\in C_0\), assume smooth
geometric and source fields, and take the patches \eqref{eq:patch}
with \(r=O(L)\).  Let \(c\) be constant on each bundle and bounded
in the coordinate neighborhood; it may depend on \(L\).
Using \eqref{eq:xic} with the original correction
\(\zeta=O(\kappa\rho^3/\cR_c^2)\), the defect obeys
\begin{equation}
 \mathfrak D^{(c)}[\cH_{L,r}]
 =\frac{A_r}{4G}L\left(\frac L2-c\right)
       (R_{kk}-8\pi GT_{kk})(p)
  +o\bigl(A_rL\max(L,|c|)\bigr).
 \label{eq:affineshrinking}
\end{equation}
Suppose the matched weighted balance is imposed to the accuracy
\(\mathfrak D^{(c)}=o(A_rL\max(L,|c|))\), and for some fixed
\(\varepsilon_0>0\),
\[
 \frac{|L/2-c|}{\max(L,|c|)}\geq\varepsilon_0
\]
along the shrinking sequence.  Then \(R_{kk}=8\pi GT_{kk}\) at \(p\).
Imposing these assumptions in every null direction at every point,
with stress-tensor conservation, yields the Einstein equation.
\end{proposition}

\begin{proof}
The defect is the integral of
\((V-c)[(R_{kk}-8\pi GT_{kk})+\sigma^2-\theta^2/2]/(4G)\)
with the physical area measure, minus
\(\int T_{ab}\zeta^ak^b\dd V\dd A/T_{\rm U}\).
Continuity, the equilibrium optical data at \(p\), and
\(\int_0^L(V-c)\dd V=L(L/2-c)\) give the displayed leading term.
The last integral is \(O(A_rL^4/\cR_c^2)\) for bounded source
components and is subleading to the stated scale.
The nondegeneracy condition permits division by the leading
coefficient.  The null-vector and conservation argument then proceeds
as in Theorem~\ref{thm:Einstein}.
\end{proof}

This includes fixed \(c\) and \(c=\gamma L\) with
\(\gamma\neq1/2\), as statements about the prescribed weighted
balance.  It does not remove the physical assumptions of
Theorem~\ref{thm:Einstein}.  At the midpoint,
\begin{equation}
 \int_0^L(V-L/2)\dd V=0,\qquad
 \int_0^L(V-L/2)V\dd V=\frac{L^3}{12}.
 \label{eq:midpoint}
\end{equation}
The leading constant-field term is therefore invisible to this
weight.  An \(o(A_rL^2)\) balance alone supplies no derivative
constraint.  Such a constraint would require a stronger
\(o(A_rL^3)\) balance and a controlled higher-order expansion;
neither is assumed here.

The exact generator-normalized equivalence in
Theorem~\ref{thm:balanceequivalence} has a simpler shift property:
replace every weight \(V\) by \(V-c\) and impose the balance on
\(c<V<c+V_*\), for every sufficiently small bundle and null
direction.  The nonzero prefactor again gives the null-projected
field equation.  The bundle must extend throughout this interval
without a caustic.  This algebraic equivalence does not depend on
the midpoint exception of the integrated shrinking-patch argument.

Thus the field-equation condition obtained from the \emph{matched}
pair \((S_{\loc}^{(c)},\xi_c)\) is unchanged under the nondegenerate
weighted-balance assumptions; the pair itself is not invariant.
Keeping \(\xi=\xi_0\) while shifting only the entropy gives the
different defect \(\mathfrak D^{(0)}+c\Delta\Pi\), whose leading
curvature and matter weights generally no longer match.

About a stationary vacuum Killing-horizon background, for fixed
anchor and source perturbations obeying the linearized Einstein
equation, the optical squares have no first variation.  Equation
\eqref{eq:affineshiftdefect} then gives
\[
 \delta(\Delta\Pi+2\pi\Delta P)
 =-\frac1{4G}\int_{\cH}
       \delta(R_{kk}-8\pi GT_{kk})\,\dd V\,\dd A_0=0.
\]
Hence the first-order defect is unchanged.  The entropy values
themselves need not agree:
\(\delta S_{\loc}^{(c)}=\delta S_{\loc}+c\,\delta\Pi\),
where \(\delta\Pi=(4G)^{-1}\int\delta\theta\,\dd A_0\).
Equality holds, in particular, on cuts with \(\delta\theta=0\).

\paragraph{Positivity and the Killing-horizon Hessian.}
Under null convergence, \eqref{eq:nonlinearc} gives
\(\dd S_{\rm L}^{(c)}/\dd V\geq0\) for \(V\geq c\), and
\(\dd S_{\rm L}^{(c)}/\dd V\leq0\) for \(V\leq c\), while the
bundle remains caustic-free.  In the stationary-vacuum setting of
Theorem~\ref{thm:secondorder},
\begin{equation}
 \Delta\delta^2S_{\loc}^{(c)}
 =\frac1{2G}\int_{\cH_{12}}(V-c)\delta\sigma^2\,\dd V\,\dd A_0
 =\Delta\delta^2S_{\loc}
   -\frac{c}{2G}\int_{\cH_{12}}\delta\sigma^2\,\dd V\,\dd A_0 .
 \label{eq:affineHessian}
\end{equation}
This follows from
\(\Delta\delta^2\Pi=-(2G)^{-1}\int\delta\sigma^2\dd V\dd A_0\).
For the fixed Killing field \(\xi\) in that theorem, the
canonical-energy identity is
\(T_{\rm U}\Delta\delta^2S_{\loc}=\int e_G[\xi;h,h]\).
Reanchoring adds the explicitly displayed shear term, multiplied by
\(T_{\rm U}\); it does not leave the identity with the same
canonical-energy flux unchanged for general radiative perturbations.
The shifted entropy Hessian can be negative on segments containing
shear to the past of \(C_c\).

On a planar Minkowski Rindler horizon, the shifted weight extends
to another exact Killing boost \(\xi_c\).  If the hypotheses of
the stationary-horizon theorem are imposed for that boost, then
\(T_{\rm U}\Delta\delta^2S_{\loc}^{(c)}
=\int e_G[\xi_c;h,h]\), nonnegative on its future branch.
This comparison changes the Killing field used to define canonical
energy.  It is not a general reanchoring symmetry on a curved
Killing horizon.  The matter Hessian and storage formulas likewise
retain their geometric form under \(V\to V-c\); for example,
\[
 \cB_{\rm exp}^{(c)}
 =\frac{\kappa}{8\pi G}\int
       \bigl(a^2-2(V-c)ay\bigr)\dd A_0 .
\]

\paragraph{Physical anchoring.}
The equilibrium cut supplies the anchor in the conditional local
Clausius construction.  The background bifurcation surface supplies
it in the exact HWZ comparison.  Reanchoring changes the weighted
functionals and fluxes; when a suitable boost interpretation exists,
it also changes the observer used to define boost energy.  A passive
coordinate change does neither.  The polarizations \(c\Pi\) and
\(c\Pi_{\rm L}\) are analogous to the change of a boost quantity
under a translation of its origin, but an identification with a
Hamiltonian translation charge or with the schematic boost
polarization of Sec.~\ref{sec:opensystem} is not made here.
Table~\ref{tab:affine} summarizes the distinctions.

\begin{table}[!htbp]
\centering
\fontsize{9.5}{11}\selectfont
\setlength{\tabcolsep}{4pt}
\setlength{\abovecaptionskip}{5pt}
\begin{tabularx}{\textwidth}{@{}>{\raggedright\arraybackslash}p{0.32\textwidth}>{\raggedright\arraybackslash}p{0.12\textwidth}>{\raggedright\arraybackslash}X@{}}
\hline
Quantity & Rescaling & Physical reanchoring to \(C_c\)\\
\hline
\(S_{\loc},S_{\rm L}\), fixed cells & invariant &
\(+c\Pi,+c\Pi_{\rm L}\), cut polarizations\\
\(S_{\rm L}\), reference measure moved & invariant &
also changes cell weights, Eq.~\eqref{eq:refchange}\\
Weighted source \(\Delta Q_{\matt}\) & invariant &
\(-c\kappa\Delta P\), Eq.~\eqref{eq:affineshiftQ}\\
Rate identities & covariant &
form-covariant with \(V\to V-c\)\\
Integrated on-shell balances & invariant &
retain their form; individual terms change\\
Local field-equation implication & invariant &
unchanged under the hypotheses of
Proposition~\ref{prop:affineEinstein}\\
Exact generator-normalized equivalence & invariant &
holds on a pre-caustic interval with \(V-c>0\)\\
First-order defect, stationary vacuum background & invariant &
unchanged for linearized on-shell perturbations\\
First-order entropy value & invariant &
\(+c\,\delta\Pi\); need not be unchanged\\
Guaranteed nondecreasing domain of \(S_{\rm L}\) & invariant &
\(V\geq c\), under null convergence\\
Entropy Hessian change & invariant &
shift given by Eq.~\eqref{eq:affineHessian}\\
HWZ energy-flux identification & invariant &
\(T_{\rm U}\Delta\delta^2S=\int e_G\) with the specified
Killing field and its bifurcation anchor\\
\hline
\end{tabularx}
\caption{Behaviour on the same geometric cuts.  Rescaling parameters
are independent of perturbations.  Physical reanchoring differs from
a passive coordinate relabeling; arbitrary shifted source weights
are not automatically physical heat.}
\label{tab:affine}
\end{table}

\FloatBarrier

\section{Pointwise local limit and open-system bookkeeping}
\label{sec:opensystem}

\subsection{What is controlled in the shrinking limit}

The exact field-equation and nonlinear identities require only null
geometry.  Their thermodynamic interpretation on a generic curved
patch also uses the approximate boost
\(\xi^a=\kappa Vk^a+\zeta^a\).  Let
\begin{equation}
 \cH_{L,r}:=
 \{(V,x^A):0\leq V\leq L,\ x\in\cD_r\subset C_0\},
 \qquad
 A_r:=\operatorname{Area}(\cD_r).
 \label{eq:patch}
\end{equation}
For \(r=O(L)\), continuity of a scalar horizon density \(F\) gives the
elementary pointwise limit
\begin{equation}
 \frac{1}{A_rL^2}
 \int_{\cH_{L,r}}\kappa V F\,\dd V\,\dd A
 =\frac{\kappa}{2}F(p)+o(1).
 \label{eq:pointwiseshrinkinglimit}
\end{equation}
Together with \(\zeta^a=O(\kappa\rho^3/\cR_c^2)\), this is sufficient
for the leading central-generator matter balance used in
Theorem~\ref{thm:Einstein}.  It does not prove a uniform
second-variation estimate for optical defects or for the non-Killing
Iyer--Wald current \cite{IyerWald}.  Such an estimate requires specified perturbation
norms, anchoring conditions, and boundary control and is not claimed
in this focused paper.  The exact second-order theorems above instead
refer to the stationary Killing-horizon setting stated explicitly in
Theorems~\ref{thm:secondorder} and \ref{thm:matterhessian}.

\subsection{Heat, radiative flux, and kinematic focusing}

The complete physical matter heat, its intrinsic fixed-cell part, and
its dilation part are
\begin{align}
 \cQ_{\rm phys}^{(2)}
 &:=
 \int_{\cH_{12}}\kappa V
 \left(\delta^2T_{kk}+2a\,\delta T_{kk}\right)
 \dd V\,\dd A_0,
 \label{eq:secondphysicalheat}\\
 \cQ_{\rm src}^{(2)}
 &:=
 \int_{\cH_{12}}\kappa V\,\delta^2T_{kk}
 \dd V\,\dd A_0,
 \label{eq:secondsourceheat}\\
 \cQ_{\rm dil}^{(2)}
 &:=
 \int_{\cH_{12}}2\kappa V a\,\delta T_{kk}
 \dd V\,\dd A_0,
 \qquad
 \cQ_{\rm phys}^{(2)}
 =\cQ_{\rm src}^{(2)}+\cQ_{\rm dil}^{(2)} .
 \label{eq:integrateddilationheat}
\end{align}
The renormalized bulk response further contains
\begin{align}
 \cD_{\rm rad}^{(2)}
 &:=
 \int_{\cH_{12}}
 \frac{\kappa V}{4\pi G}\delta\sigma^2
 \,\dd V\,\dd A_0,
 \label{eq:radiativeproduction}\\
 \cK_{\rm foc}^{(2)}
 &:=
 \int_{\cH_{12}}
 \frac{\kappa V}{8\pi G}(\delta\theta)^2
 \,\dd V\,\dd A_0,
 \label{eq:kinematicfocusing}\\
 \Pi_{\rm ren}^{(2)}
 &:=\cQ_{\rm src}^{(2)}
 +\cD_{\rm rad}^{(2)}+\cK_{\rm foc}^{(2)}.
 \label{eq:bulkentropychange}
\end{align}
Here \(\cQ_{\rm phys}^{(2)}\) is externally supplied heat, whereas
\(\cQ_{\rm src}^{(2)}\) is its intrinsic part per fixed background
cell.  The difference \(\cQ_{\rm dil}^{(2)}\) is reversible area-strain
bookkeeping and is absorbed into focusing plus endpoint storage in
\eqref{eq:renormalizedmatterbalance}.  Thus
\(\Pi_{\rm ren}^{(2)}\), not \(\cQ_{\rm src}^{(2)}\) alone, is the
renormalized matter--gravity response.  The shear square is the HWZ
gravitational canonical-energy flux under the hypotheses of
Theorem~\ref{thm:secondorder}; it is the radiative gravitational
flux term.  The expansion square is nonradiative,
kinematic focusing.  Interpreting it as a coarse-grained entropy
contribution would require justification of the additional postulate
in Sec.~\ref{sec:physicalentropy}; causal restriction alone does not
establish that interpretation.

For a general source neither \(\cQ_{\rm src}^{(2)}\) nor
\(\cQ_{\rm phys}^{(2)}\) has a prescribed sign.  For a differentiable
two-sided family satisfying the null energy condition about vacuum,
\(\delta T_{kk}=0\) and \(\delta^2T_{kk}\geq0\).  Hence
\(\cQ_{\rm dil}^{(2)}=0\),
\(\cQ_{\rm phys}^{(2)}=\cQ_{\rm src}^{(2)}\), and all three terms in
\(\Pi_{\rm ren}^{(2)}\) are nonnegative.

\subsection{Schematic finite-subsystem extension}

Comparing a null cut with a nearby marginally outer trapped surface
\(\widetilde C\) introduces additional endpoint data.  The boost
polarization
\(\mathfrak P_{\cD,\mathrm{boost}}^{(2)}\) is the cut charge associated
with varying the normalized horizon generator.  The marginal-surface
polarization \(\mathfrak P_{\rm MOTS}^{(2)}\) is the quadratic
stability energy of the normal displacement, supplemented by the
normal Green flux through the rim.  They are supported on different
boundary strata and cannot be identified or discarded
\cite{AnderssonMarsSimon}.  Boundary and corner data of this kind are
the subject of the covariant Hamiltonian treatment of spatially
bounded regions \cite{AncoTungProperties,AncoTungBoundary} and of the
local-subsystem and edge-mode analyses of the covariant phase space
\cite{DonnellyFreidel,SperanzaPhase}.

With a specified horizon-locking gauge, MOTS stability operator,
adjoint weight, displacement boundary conditions, and rim Green
current, the expected open-system bookkeeping has the form
\begin{equation}
 \frac{T_{\rm U}}{4G}
 \Delta\delta^2A[\widetilde C]
 =
 \Pi_{\rm ren}^{(2)}
 +\Delta\mathfrak P_{\cD,\mathrm{boost}}^{(2)}
 +\Delta\mathfrak P_{\rm MOTS}^{(2)}
 \qquad\text{(schematic)}.
 \label{eq:schematicareabalance}
\end{equation}
Equation \eqref{eq:schematicareabalance} is not a theorem of the
present manuscript: the required operator definitions and Green
identity are deliberately not reproduced in this physics-focused
version.  It records which terms a complete finite-subsystem theorem
must contain and prevents the bulk entropy balance from being
mistaken for an unconditional marginal-area law.  Positivity of the
bulk terms would imply area increase only after the boost
normalization is fixed, the rim is insulated, the initial
polarization is controlled, and the final weighted MOTS quadratic
form is nonnegative.  The same second-order heat also admits a
quasilocal Hamiltonian surface-flux representative
\cite{TungYuFlux}, whose relation to the present bulk balance we do
not develop here.

\section{Discussion}

The construction has three logical levels.  First, given the inputs
listed in Theorem~\ref{thm:Einstein}, the cut-local HWZ representative
produces an equation of state in Jacobson's conditional sense:
\[
 \text{matter boost heat}
 =
 T_{\rm U}\times\text{local entropy change}.
\]
Requiring this balance in equilibrium for every null direction yields
the Einstein equation.  Neither the gravitational symplectic current
nor canonical energy enters this derivation.

Second, the generator-relative functional gives the exact nonlinear
bookkeeping identity
\[
\begin{aligned}
 &\text{instantaneous-area boost-weighted matter flux}\\
 &\quad+\text{reference-measure conversion remainder}\\
 &\qquad+\text{positive optical term}\\
 &=T_{\rm U}\times\text{change of the generator-relative functional}.
\end{aligned}
\]
The conversion is required because instantaneous-area flux is
integrated with \(J\dd A_{\rm ref}\), whereas the exact positive
identity uses fixed generator cells.  It is not an extra energy
exchange: at second order it subtracts precisely the dilation heat
\(2\kappa V\int a\,\delta T_{kk}\dd A_{\rm ref}\) caused by changing
the collecting area of a cell carrying first-order flux.  The
reference data are \(C_{\rm ref}\) and its cells; the identity remains
valid until caustic formation.  Under
the causal Lagrangian coarse-graining of
Sec.~\ref{sec:physicalentropy}, \(S_{\rm L}\) is proposed as a
finite-amplitude entropy.  Because that postulate is not derived, we
do not claim that its monotonicity is an \(H\)-theorem.
Without this coarse-graining, the shear-free example
\eqref{eq:homogeneoussecondlaw} shows only a Lyapunov property of
focusing, not microscopic irreversibility.

Third, expansion about stationary equilibrium produces the HWZ
modified canonical energy.  The shear part is the radiative
gravitational flux.  The expansion square is kinematic focusing
and combines with reversible endpoint storage.  A complete open
marginal subsystem would also contain cut and rim polarizations, as
shown schematically in \eqref{eq:schematicareabalance}; no MOTS area
theorem is asserted here.

The construction does not assert that \(S_{\rm L}\) is a universal
black-hole Noether or von Neumann entropy.  It depends on
\(C_{\rm ref}\) and on the affine origin, whose change at fixed reference measure adds
the anchoring polarization of Proposition~\ref{prop:affine}.  It is defined only
while \(J>0\), and its simple positive balance uses matter energy
normalized per reference cell.  Nor is the
Unruh thermality of a finite approximate-boost patch derived here.
The physically anchored result is the equilibrium first law and
canonical-energy Hessian under the hypotheses stated above.  The
finite-amplitude result is a geometric identity equipped with a
clearly stated coarse-grained entropy postulate, whose microscopic
derivation remains open.

It is therefore worth stating plainly what is established
independently of the finite-amplitude entropy postulate.  Four results
require no such postulate and no coarse-graining; where \(T_{\rm U}\) appears, only the Unruh
normalization \(\kappa/2\pi\) is used, and no finite-patch thermality
derivation is invoked.  First, the cut-local
identity of Theorem~\ref{thm:offshell} and the generator-relative
identity of Theorem~\ref{thm:nonlinearidentity} are exact consequences
of the Raychaudhuri equation alone; the second converts the full
nonlinear focusing equation into a form in which the curvature,
shear, and expansion terms all enter with positive sign --- and which
is therefore nonnegative whenever \(R_{kk}\geq0\) --- whereas the cut-local HWZ representative
has a negative expansion-square contribution in its derivative.  Second, under the null
convergence condition \(R_{ab}k^ak^b\geq0\) --- which on Einstein
solutions follows from the null energy condition --- \(S_{\rm L}\) is a
monotone functional of every twist-free null congruence for
\(V\geq0\) up to the first caustic
(Corollary~\ref{cor:monotonefunctional}), and for \(V\geq c\) under
any other anchoring (Proposition~\ref{prop:affine}).
This is a focusing statement about null geometry and
holds whether or not \(S_{\rm L}\) is read thermodynamically.
The proposed coarse-graining in Sec.~\ref{sec:physicalentropy}
does not by itself establish an entropy interpretation.  Third, by
Theorem~\ref{thm:balanceequivalence} the generator-normalized balance,
imposed for every sufficiently small bundle and every null direction,
is \emph{equivalent} to the Einstein equation with an undetermined
cosmological constant, with no energy condition assumed; the nonlinear
identity is thus a finite-amplitude geometric recasting of the
null-projected field equation, characterizing it rather than
presupposing it.
Fourth, for stationary vacuum perturbations under the hypotheses of
Theorem~\ref{thm:secondorder} the temperature-weighted change of
the entropy Hessian \emph{equals} the HWZ modified canonical-energy
flux, so the
near-equilibrium entropy response is fixed by covariant phase space
rather than assumed; this identification requires the anchoring cut
to be the bifurcation cut of that Killing field
(Eq.~\eqref{eq:affineHessian}).  The physical content obtained here is
accordingly a single geometric functional that reproduces the Einstein
equation in equilibrium, carries the established canonical-energy
second law as its equilibrium Hessian, and remains monotone at finite
amplitude.

\section*{Acknowledgements}

The author thanks the anonymous referee for comments that improved the
clarity and precision of the manuscript.

\end{document}